\UseRawInputEncoding
\documentclass{elsarticle}

\usepackage{hyperref}             

\usepackage{IEEEtrantools}

\usepackage{graphics,graphicx}

\usepackage{amsmath}
\usepackage{amsfonts}
\usepackage{mathtools}
\usepackage{tabularx}
\usepackage{booktabs}	
\usepackage{siunitx}
\usepackage{textcomp}
\usepackage{placeins}

\usepackage{stfloats}

\usepackage[shortlabels]{enumitem}

\usepackage{xcolor}
\usepackage{color,soul}

\usepackage[bordercolor=white,backgroundcolor=gray!30,linecolor=black,colorinlistoftodos]{todonotes}

\usepackage{accents}

\usepackage{nomencl}

\usepackage{setspace}
\usepackage{inputenc}
\usepackage{textcomp}

\usepackage[flushleft]{threeparttable}
\usepackage{gensymb}
\usepackage{ifthen}
\renewcommand{\nomgroup}[1]{%
	\ifthenelse{\equal{#1}{P}}{\item[\newline]\item[\textit{Parameters}]}{%
		\ifthenelse{\equal{#1}{I}}{\item[\newline]\item[\textit{Indices and Superscripts}]}{%
			\ifthenelse{\equal{#1}{J}}{\item[\newline]\item[\textit{Sets}]}{%
				\ifthenelse{\equal{#1}{V}}{\item[\newline]\item[\textit{Variables}]}{
					\ifthenelse{\equal{#1}{A}}{\item[\newline]\item[\textit{Acronyms}]}{}
				}
			} 
		}
	}
}

\usepackage{multirow}
\usepackage{rotating}

\usepackage[]{tikz}
\newlength\mylinewidth
\usepackage{pgfplots}
\pgfplotsset{width=0.8\textwidth,height=2.5in,compat=1.16}

\makenomenclature

\journal{Applied Energy}

\begin{document}

\begin{frontmatter}

\title{Toward Transactive Control of Coupled Electric Power and District Heating Networks}


\author[mymainaddress]{Jona Maurer\corref{mycorrespondingauthor}}
\cortext[mycorrespondingauthor]{Corresponding author}
\ead{jona.maurer@kit.edu}

\author[mymainaddress]{Nicolai Tschuch}
\author[mymainaddress]{Stefan Krebs}
\author[mysecondaryaddress]{Kankar Bhattacharya}
\author[mysecondaryaddress]{Claudio Ca{\~n}izares}
\author[mymainaddress]{S\"oren Hohmann}

%

\address[mymainaddress]{Karlsruhe Institute of Technology, Institute of Control Systems, \\Wilhelm-Jordan-Weg 1, 76131 Karlsruhe, Germany}
\address[mysecondaryaddress]{Department of Electrical and Computer Engineering, University of Waterloo, \\ON, N2L3G1, Canada}

\begin{abstract}
Although electric power networks and district heating networks are physically coupled, they are not operated in a coordinated manner. With increasing penetration of renewable energy sources, a coordinated market-based operation of the two networks can yield significant advantages, as reduced need for grid reinforcements, by optimizing the power flows in the coupled systems. Transactive control has been developed as a promising approach based on market and control mechanisms to coordinate supply and demand in energy systems, which when applied to power systems is being referred to as transactive energy. However, this approach has not been fully investigated in the context of market-based operation of coupled electric power and district heating networks. Therefore, this paper proposes a transactive control approach to coordinate flexible producers and consumers while taking into account the operational aspects of both networks, for the benefit of all participants and considering their privacy. A nonlinear model predictive control approach is applied in this work to maximize the social welfare of both networks, taking into account system operational limits, while reducing losses and considering system dynamics and forecasted power supply and demand of inflexible producers and consumers. A subtle approximation of the operational optimization problem is used to enable the practical application of the proposed approach in real time. The presented technique is implemented, tested, and demonstrated in a realistic test system, illustrating its benefits.
\end{abstract}


\end{frontmatter}


\nomenclature[aepn]{EPN}{Electric Power Network}
\nomenclature[ac]{CEPDHN}{Coupled Electric Power and District Heating Network}
\nomenclature[ad]{DHN}{District Heating Network}
\nomenclature[ar]{RES}{Renewable Energy Source}
\nomenclature[ac]{CHP}{Combined Heat and Power}
\nomenclature[ate]{TE}{Transactive Energy}
\nomenclature[atce]{TC}{Transactive Control}
\nomenclature[atcs]{TCS}{Transactive Control System}
\nomenclature[af]{FNP}{Flexible Network Participant}
\nomenclature[ai]{ISOEMS}{Independent System Operator Energy Management System}
\nomenclature[anmpc]{NMPC}{Nonlinear Model Predictive Control}
\nomenclature[anlp]{NLP}{Nonlinear Programming}
\nomenclature[aems]{EMS}{Energy Management System}
\nomenclature[aemssa]{EMSSA}{Energy Management System Software Agent}
\nomenclature[adso]{DSO}{Distribution System Operator}
\nomenclature[adsm]{DSM}{Demand Side Management}
\nomenclature[adpr]{DPR}{Differential Pressure Regulator}
\nomenclature[am]{MINLP}{Mixed Integer Nonlinear Programming}
\nomenclature[aw]{WPP}{Wind Power Plant}
\nomenclature[ah]{HP}{Heat Pump}
\nomenclature[au]{UMP}{Uniform Marginal Price}
\nomenclature[al]{LMP}{Locational Marginal Price}
\nomenclature[ab]{BESS}{Battery Energy Storage System}
\nomenclature[ap]{PV}{Photovoltaic}
\nomenclature[vm]{$ \boldsymbol{\dot{m}}$}{Vector of mass flows through components in [$\SI{}{kg}/\SI{}{s}$]}
\nomenclature[vp]{$ \Delta\boldsymbol{p}$}{Vector of edge pressure differences in [$\SI{}{bar}$]}
\nomenclature[vp]{$P_i,~P_n$}{Real power infeed/demand in [p.u.] at a bus $i$, \\of a network participant $n$}
\nomenclature[vq]{$Q_i, Q_n$}{Reactive power infeed/demand in [p.u.] at a bus $i$, of a network participant $n$}
\nomenclature[vv]{$V_i$}{Voltage amplitude at a bus $i$ in [p.u.]}
\nomenclature[vza]{$\delta_i$}{Voltage angle at a bus $i$ in [$\SI{}{\radian}$]}
\nomenclature[vw]{$W$}{Social welfare in [\UseTextSymbol{TS1}{¤}]}
\nomenclature[vk]{$k$}{Time step}
\nomenclature[vk]{$K_{\mathrm{v}}$}{Flow coefficient of a valve in [$\SI{}{\cubic\meter}/\si{\second}$]~at~\SI{1}{\bar}}
\nomenclature[vp]{$\Delta p^{\mathrm{dpr}}$}{Pressure diff. over a differential pressure regulator}
\nomenclature[vr]{$R,~S$}{Water mass in [kg], see Figure \ref{fig_node_method}}
\nomenclature[vp]{$t_l$}{Medium time of stay in [$\si{\second}$] of water leaving pipe $l$}
\nomenclature[vt]{$T_{l}^\mathrm{out_1}$}{Lossless outlet temperature of a pipeline in [$\si{\kelvin}$]}
\nomenclature[vt]{$T_{l}^\mathrm{out_2}$}{Lossy outlet temperature of a pipeline in [$\si{\kelvin}$]}
\nomenclature[vt]{$T_{c}^\mathrm{out}$}{Outlet temperature of a component in [$\si{\kelvin}$]}
\nomenclature[vzb]{$\gamma, ~\varepsilon$}{Auxiliary integer variable to determine R and S}
\nomenclature[vzd]{$\Phi_n$}{Thermal power infeed/demand in [$\si{\mega\watt}$] at node $i$}
\nomenclature[vze]{$\omega$}{Auxiliary weighting variables for the node method}

\nomenclature[pt]{$\Delta T^{\mathrm{prod}}$}{Maximal outlet temperature change in [$\si{\kelvin}$]}
\nomenclature[pp]{$ \Delta\boldsymbol{p}^{\mathrm{def}}$}{Vector of predefined pressure differences in [$\SI{}{bar}$]}
\nomenclature[pc]{$c_\mathrm{p}$}{Specific heat capacity of water in [$\si{\joule}/(\si{\kilogram}~\si{\kelvin})$]}
\nomenclature[pa]{$\boldsymbol{A}$}{Edge-node incidence matrix of the DHN}
\nomenclature[pb]{$\boldsymbol{B}$}{Edge-loop incidence matrix of the DHN}
\nomenclature[pr]{$\boldsymbol{R}$}{Edge control path matrix of the DHN}
\nomenclature[pb]{$B_{ij}$}{Susceptance of distribution feeder in [p.u.]}
\nomenclature[pg]{$G_{ij}$}{Conductance of distribution feeder in [p.u.]}
\nomenclature[pn]{$I,~N$}{Number of network nodes/buses $i$, participants $n$}
\nomenclature[pc]{$c$}{Bid price [\UseTextSymbol{TS1}{¤}/$\si{\mega\watt}$]}
\nomenclature[pp]{$\Delta p^{\mathrm{pump}}$}{Pressure difference over a controlled pump in [$\si{\bar}$]}
\nomenclature[pp]{$\Delta p_{0}$}{Pressure reference for valve flow factor $K_{\mathrm{v}}$ in [$\SI{}{\bar}$]}
\nomenclature[pzc]{$\rho$}{Density of water in [$\si{\kilogram}/\si{\cubic\meter}$]}
\nomenclature[pk]{$\Delta k$}{Time intervall between two time steps in [$\si{\second}$]}
\nomenclature[pl]{$L$}{Length of a pipeline in [$\si{\meter}$]}
\nomenclature[pa]{$A_l$}{Cross section area of a pipeline in [$\si{\square\meter}$]}
\nomenclature[pr]{$R'_l$}{Thermal resistance of a pipeline in [$(\si{\meter}~\si{\kelvin})/\si{\watt}$]}
\nomenclature[pt]{$T^\mathrm{a}$}{Ambient temperature in [$\si{\kelvin}$]}
\nomenclature[pza]{$\zeta_n$}{Coupling factor of an energy converter $n$}
\nomenclature[pzb]{$\mu$}{Friction factor in [$(\si{\bar}~\si{\second}^2)/\si{\kilogram}^2$]}
\nomenclature[pzc]{$\_,^{-}$}{Minimum and maximum values}
\nomenclature[ic ]{$\mathrm{c,~p}$}{Consumer, Producer}
\nomenclature[ii ]{$i,~j$}{Buses/Nodes $i,j=1,...,I$}
\nomenclature[in ]{$n,~m$}{Network participants $n,m=1,...,N$}
\nomenclature[icp ]{$\mathrm{cp}$}{Control path}
\nomenclature[il ]{$l$}{Pipeline}
\nomenclature[iza ]{$0$}{Reference value for valve flow factor $K_{\mathrm{v}}$}
\nomenclature[ie ]{$e$}{Edge/Component of the DHN}
\nomenclature[ir ]{$\mathrm{r,~s}$}{Return/Supply network of the DHN}
\nomenclature[izb ]{$^\mathrm{*}$}{Predicted power infeed/demand}
\nomenclature[ie ]{$\mathrm{el,~ht}$}{Electric, Heat}


\printnomenclature[0.7in]

\section{Introduction}

The challenging task of transforming the conventional energy systems into integrated energy systems, and its associated benefits are well known \cite{WuYanJiaEtAl2016,Mancarella2014}. However, due to the increasing integration of Renewable Energy Sources (RESs) into Electric Power Networks (EPNs), and the requirement to balance power supply and demand at all times, the need for flexibility and integration of EPNs, with other energy systems is become quite relevant. This has led to extensive research into new forms of grid operation for EPNs, to enable temporal and spatial coordination of Flexible Network Participants (FNPs). In this context, the concepts of Transactive Energy (TE) and Transactive Control (TC) have been developed \cite{KokWidergren2016}. Since multiple definitions of TE and TC exist, in this paper, TE is defined based on the GridWise Architecture Council definition \cite{AbrishambafLezamaFariaEtAl2019,Council2015}: ``a system of economic and control mechanisms that allows the dynamic balance of supply and demand across the entire electrical infrastructure using value as a key operational parameter." Furthermore, the following concept of TC, as defined in \cite{LiLianJ.ConejoEtAl2020}, is used here: ``a domain-free approach that integrates market-based coordination and value-based control for a group of resources to achieve certain global objectives."  In this context, TC enables optimal coordination of FNPs while overcoming the disadvantages of earlier concepts such as direct load control and price-responsive control, as the former does not consider user preferences and the latter lacks predictability of the load response. When TC is applied to an EPN, it is usually referred to, as a TE system \cite{LiLianJ.ConejoEtAl2020}.

A wide range of papers has been published in the field of TE. For example an approach based on agents managing energy transactions of bordering microgrids, in the context of network connections is presented in \cite{JankoJohnson2018}. An Energy Management System (EMS), based on a decomposition of the main problem into multiple sub-problems for different types of facilities and their coordination is proposed in \cite{AkterMahmudHaqueEtAl2020}. An overview of the wide field of research in the context of TE for EPNs is presented in \cite{AbrishambafLezamaFariaEtAl2019,LiLianJ.ConejoEtAl2020,LiuWuHuangEtAl2017}.

Driven by the need to achieve full decarbonization of heat supply, District Heating Network (DHN) operators are now facing similar challenges as EPN operators. Thus, there are a rising number of FNPs in DHNs needing coordination to achieve safe and efficient network operation, which requires new operating strategies. Furthermore, the coupling between DHNs and EPNs is increasing \cite{WuYanJiaEtAl2016}, with the integration of Combined Heat and Power (CHP) plants, Heat Pumps (HPs) and Electric Boilers (EBs), which requires coordination between heat and power systems. However, currently, EPNs and DHNs are operated separately, with some DHN operators participating in electricity markets as suppliers and/or consumers, which presents various techno-economic issues that have been discussed in the existing literature. Thus, while the price of heat in DHNs is based on electricity prices, which results in cross-subsidies \cite{DengLiSunEtAl2019}, on the other hand, the benefits of using DHNs as sources of flexibilities for the EPN, with their energy storage potential in pipelines, heat storage units such as large water tanks, and the thermal storage capacity of buildings \cite{VandermeulenHeijdeHelsen2018}, are not being fully utilized. In this context, there is a critical need to design a TC approach for Coupled Electric Power and District Heating Networks (CEPDHNs), so that both networks can be optimally operated, while considering FNPs.

Due to the fundamental difference of the underlying physics, the resulting models, and the implementation of network operation of EPNs and DHNs, it is not possible to apply TC approaches developed for EPN operation to DHNs. These differences are the relatively slow propagation of heat and the concomitant potential to store energy within DHNs. Furthermore, flow directions are determined by set points of valves and pumps, and are not dependent on the power infeed or demand of FNPs in DHNs. Also, DHN models need to incorporate supply and return networks, while directional power flows can be modeled on a feeder section of a distribution system in EPNs. Additionally, DHN potentials are represented by two variables, pressure and temperature, while EPN potentials are defined by the voltage magnitude. Finally, reactive power flows are only found in EPNs.

Given the above differences, TC approaches for CEPDHNs have not been fully studied, and the application of TC techniques to multi-energy systems is relatively new. For example, a TE modeling and assessment framework for distributed multi-energy systems is presented in \cite{GoodCesenaHeltorpEtAl2018}, and an alternating direction method of multipliers is used in a TE system to coordinate the distributed energy sharing among multi-energy microgrids in \cite{YangHuAiEtAl2020}. The authors of \cite{WangJiaHouEtAl2019} propose a double auction retail market framework to enable optimal supply and consumption with electricity and heat. In \cite{BehboodiChassinDjilaliEtAl2018}, thermostatic loads of heat pumps and air conditioners are operated based on the TC paradigm, while maintaining comfortable temperatures. A double stage stochastic approach, which allows thermal energy storages of buildings to manage the uncertainty resulting from energy procurement in a TC context, is reported in \cite{YuPavlak2021}. Finally, a TE approach based on peer-to-peer transactions in the context of multiple energy hubs, with multiple energy carriers and high infeed of RES, is discussed in \cite{GanYanYaoEtAl2021}. However, none of these papers discusses the following relevant issues for TC of CEPDHNs: 
\begin{itemize}
	\item Combining market and control mechanisms: Spatial and temporal coordination of power infeed and demand of all FNPs, such as producers, consumers, storages, and energy converters, should be achieved by a combined implementation of market and control mechanisms in a TC approach to achieve safe and efficient operation of the CEPDHN.
	\item Flexibility preserving DHN operation with low losses: Heat losses mainly occur in DHNs, based on the heat flow through pipeline insulations. As heat losses increase with rising difference between the fluid and the ambient temperature, these can be minimized by keeping the temperatures as low as possible \cite{ZhangLiGudmundsson2013}\footnote{Supply network temperatures can only be reduced to a certain level in order to sufficiently supply the heat demand of customers.}. Therefore, DHN operators aim to set the mass flows through the heat exchangers of FNPs such that the temperature difference between supply and return network is maximized, and thereby the mass flows and the resulting pump costs are minimized \cite{ZhangLiGudmundsson2013}. These mass flows are adjusted by pumps, Differential Pressure Regulators (DPRs), their respective control paths, and valves. As heat losses are in the range of $\SI{12}{\percent}\text{~to~}\SI{20}{\percent}$ \cite{VesterlundSandbergLindblomEtAl2013,ComakliYuekselComakli2004,LundMoellerMathiesenEtAl2010}, an efficient operation based on variable mass flows and variable temperatures is of significant interest. Furthermore, by storing thermal energy in the transported fluid, DHNs can provide flexibility to the CEPDHN.
	\item Limiting computational costs: The complexity of the network models used in the CEPDHN optimization problem associated with a TC approach may result in significant computational costs. However, in practice, these mathematical models should be solved in real time. This is especially important for DHN TC models, since such models do not exist in the current literature.
	\item Preserving privacy for FNPs \cite{KokWidergren2016,LiLianJ.ConejoEtAl2020}: Based on smart meter data, sensitive information of FNPs can be collected \cite{McKennaRichardsonThomson2012}. Hence, a well designed TC approach should preserve the privacy of the facilities in a CEPDHN. 
\end{itemize}
Based on the aforementioned discussions, this paper proposes a new TC approach for CEPDHNs as follows: For the optimal economic operation of CEPDHNs, FNPs should be operated in a way that maximizes social welfare. To this effect, an intra-day Transactive Control System (TCS)  auction market, activated after settlement of a day-ahead energy market, is assumed, to coordinate the interests of all FNPs. The proposed intra-day energy market is auction based and considered within an Independent System Operator EMS (ISOEMS), using a Nonlinear Model Predictive Control (NMPC) approach \cite{Gruene2017}. Therefore, the ISOEMS takes into account the nonlinear dynamic and stationary network models, operational constraints, predictions of power supply and demand of inflexible units, and bids of FNPs to determine the optimal control values that maximizes social welfare and ensures secure network operations at the same time. Thereby, the proposed market and control mechanisms are integrated into a new TCS to optimally operate the CEPDHNs. Hence, the main contributions of the presented work are:
\begin{itemize}
	\item The proposed novel auction-based TC approach presents a technically efficient market-based network operation, considering privacy issues of the FNPs \cite{KokWidergren2016}, while communicating their bids/offers without passing technical information such as indoor temperatures and state of charge of batteries to other entities as proposed by earlier load control approaches \cite{LiLianJ.ConejoEtAl2020}. 
	\item A detailed thermo-hydraulic DHN model is developed within the ISOEMS, which considers the pressure differences and mass flows caused by varying consumer behavior, which are important in the context of Demand Side Management (DSM), while incorporating heat propagation through pipelines with variable temperatures and mass flows, enabling the representation of pipeline storage. These allow the EPN to profit from the full flexibility of the DHN, which has also been noted in \cite{ZhangWuWenEtAl2021}, where future work for CEPDHNs is identified as: ``modeling for schemes promoting the participation of flexible devices can be explored to further increase the flexibility of the system''.
	\item A computationally efficient CEPDHN Nonlinear Programming (NLP) model is proposed, which considers the impact of active controlled hydraulic components such as DPRs, valves, and pumps, and their mass flows and differential pressures throughout the DHN, based on an approximated but accurate pipeline model, allowing to incorporate the most efficient form of DHN operation in the proposed TC approach. This is of importance as CEPDHN models easily result in high computational costs \cite{ZhangWuWenEtAl2021}. 
\end{itemize}
It is important to highlight that TE applications for EPNs have focused on distributed control/optimization approaches, while the proposed TC approach is based on a centralized optimization model. This is because a system operators point of view is considered in this paper, as opposed to the market participants', like other existing studies, since it fits better in existing EPNs and likely future DHN market structures, which are mostly centralized, and where FPNs are treated as providers of flexibility in a more traditional DSM context rather than peer-to-peer approaches. Furthermore, the presented centralized approach, allows to better understand the proposed merged market-based operation of CEPDHNs, which is not yet well understood, thus obtaining reference conditions and parameters for future studies of decentralized operation of such integrated networks. Finally, the proposed approach is closer to the way EPNs and DHNs are operated currently, and hence more likely to be implemented in practical applications, as it unites the functionality of auction platforms, bidding agents, and EMS.
\addtocounter{footnote}{-1}  
\begin{figure}
	\centering
	\includegraphics[width=0.8\textwidth]{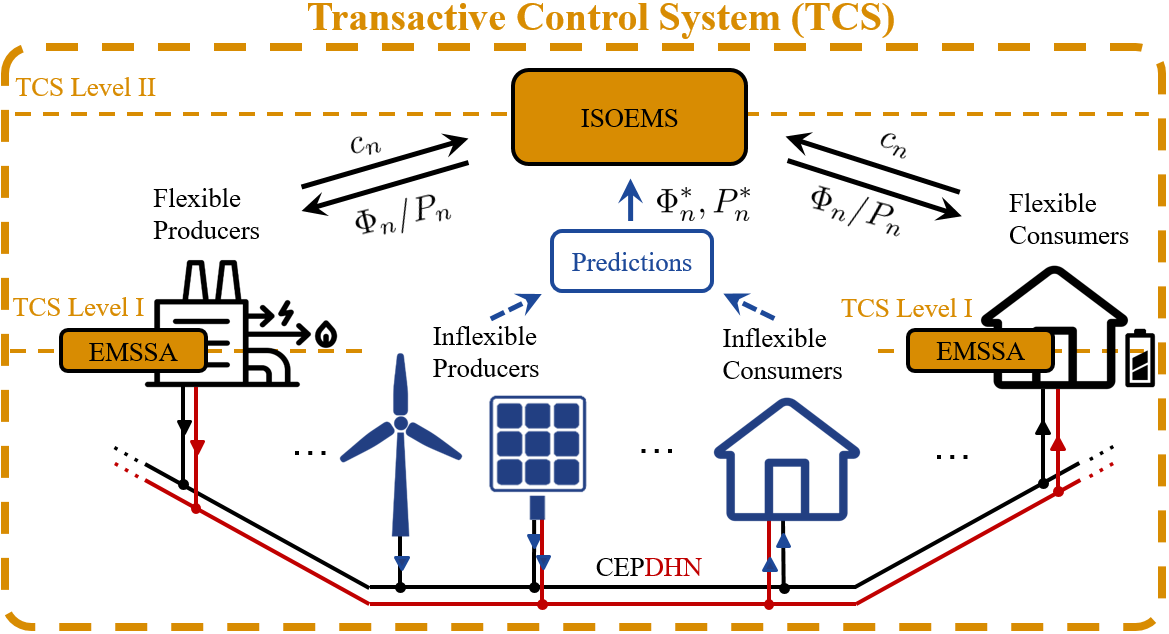}
	\caption{Proposed approach showing information flow between the two hierarchy levels of the TCS, i.e., the ISOEMS (TCS Level II) and the EMSSAs of the FNPs (TCS Level I), and the power flow between four different groups of network participants connected to the CEPDHN\protect\footnotemark.}
	\label{fig_approach}
\end{figure}
\footnotetext{Credit for initial images of cogeneration, wind turbine, solar panel, home, and battery goes to \cite{RinaldiK.KimDomingoEtAl}.}

The rest of the paper is organized as follows: The proposed TCS approach and model are presented in Section 2. The results for realistic studies demonstrating the effectiveness of the proposed TCS are discussed in Section 3. Finally, Section 4 highlights the main conclusions and contributions of the paper.

\section{Transactive Control System}
\label{sec:TCS}

An overview of the proposed TCS is shown in Figure \ref{fig_approach}, which can be described as follows, based on the nomenclature defined at the beginning of the paper:
\begin{enumerate}
	\item Every EMS Software Agent (EMSSA) of an FNP $n$ taking part in the TCS auction market sends its bids/offers $c_{n}$ to the ISOEMS. Additionally, predictions of further power infeed $P^*_n$ and demand $\Phi^*_n$, of inflexible network participants are sent to the ISOEMS by the network operators.  
	\item The ISOEMS calculates the optimal control values $P_n$ and $\Phi_n$ for all FNPs maximizing the social welfare $W$, while taking into account the operational limits of the CEPDHN. 
\end{enumerate}
This procedure is repeated for several time steps, based on a rolling horizon approach of the NMPC \cite{Gruene2017}. The bids/offers, predictions, and control values are sent and determined for all time steps of the prediction horizon\footnote{FNPs are incentivized to keep bids/offers for a specific time step constant over the prediction horizon to avoid penalty fees. However, the proposed approach can also handle cases where FNPs change their bids/offers over time.}.

The proposed TCS auction market is distinctly different from the traditional wholesale energy markets, where the participating agents are conventional generators and loads with bulk bids/offers, the market prices may be Locational Marginal Prices (LMPs) or Uniform Marginal Prices (UMPs), and the operational time frame can range from real-time, to hour-ahead, intra-day, day-ahead, forward markets, and involves a larger jurisdiction such as a province/state or even a country. On the other hand, the proposed TCS auction is a multi-period combined auction of two coupled market platforms associated with the EPN and DHN, respectively, where both markets are cleared in a coordinated manner and simultaneously, which is the core difference with standard auction market frameworks.

The energy dispatches in the proposed TCS auction market, operating in 15-minute intervals, change for every time step using an NMPC technique, in order to optimally adapt to the updated forecasts of supply and demand of inflexible network participants. The bids and offers of the FNPs are assumed to be provided by local EMSSAs, such as existing facility EMSs. These automatically send bids to the ISOEMS for each time step based on the calculated flexibility of their respective facility, such as offices, factories, Battery Energy Storage Systems (BESS), and/or energy hubs. Observe that these bids and offers are only known to the individual market participant and the TCS auction market operator, thus enhancing privacy as defined in \cite{KokWidergren2016}. No other information such as indoor temperatures, state of charge of batteries, or electric vehicle data is shared. Note that privacy is a direct outcome from the information sharing specifications in the proposed TCS. Other approaches, such as direct load control \cite{LiLianJ.ConejoEtAl2020}, which coordinate flexible resources based on monetary incentives, require sharing information not needed in the presented TCS.

The flexibility of consumer facilities can be determined by their EMSSAs by first calculating the predicted minimum power demand for the next hours and the expected usable amount of energy \cite{alrumayh2019flexibility}. Based on these, two profile bids with minimum and maximum power values can be calculated. The bid prices of the FNPs are determined by their respective EMSSAs using an approach such as \cite{adika2014demand}. 

In Germany, today the Distribution System Operator (DSO) by itself, operates the electric power, gas, water, and district heating systems. For jurisdictions where there are multiple entities responsible for utility services of electricity and heat, there has to be information sharing agreements between the entities in order to implement the proposed framework and models. This will require some coordination and policy formation at the regulatory/municipal governmental levels. For the sake of generality, it has been assumed that such information sharing and privacy protection mechanisms are already in place in such jurisdictions.

\subsection{ISOEMS Objective Function}
The welfare function $W$ of the CEPDHN is formulated as the sum of the benefits of the $N_\mathrm{c}^\mathrm{el}$ EPN and $N_\mathrm{c}^\mathrm{ht}$ DHN consumers, net costs of $N_\mathrm{p}^\mathrm{el}$ EPN and $N_\mathrm{p}^\mathrm{ht}$ DHN producers, for all time intervals $k$, given as follows:
\begin{IEEEeqnarray}{rClCl}
	W_k =& \sum_{n=1}^{N_{\mathrm{c}}^{\mathrm{el}}}c_{n,k} P_{n,k} - \sum_{n=N_{\mathrm{c}}^{\mathrm{el}}+1}^{N^\mathrm{{el}}=N_{\mathrm{c}}^{\mathrm{el}}+N_{\mathrm{p}}^{\mathrm{el}}}c_{n,k} P_{n,k} \nonumber \\
	&+\sum_{n=N^{\mathrm{el}}+1}^{N^{\mathrm{el}}+N_{\mathrm{c}}^{\mathrm{ht}}}c_{n,k} \Phi_{n,k} - 
	\sum_{n=N^{\mathrm{el}}+N_{\mathrm{c}}^{\mathrm{ht}}+1}^{N^{\mathrm{el}}+N_{\mathrm{c}}^{\mathrm{ht}}+N_{\mathrm{p}}^{\mathrm{ht}}}c_{n,k} \Phi_{n,k}
	\label{eq:SW}
\end{IEEEeqnarray}

Here, electric storage units may appear either as a consumer (when charging) or as producer (when discharging), and inflexible participants are considered to provide a zero-priced bid/offer, $c_{n,k}=0$. 
\subsection{CEPDHN Model for ISOEMS}

The CEPDHN model comprises three main modules: the first corresponds to the model of the EPN, the second to the energy converters, and the third to the DHN model. DHNs are typically modeled by a stationary hydraulic model and a dynamic thermal model, due to the large difference in the propagation speed of hydraulic effects, which travel at the speed of sound, and the maximum flow velocity of water in DHNs, which is around \SI{3}{\meter/\second} \cite{Oppelt2015}. Therefore, the propagation of thermal fronts has to be described in the model for controlling the CEPDHN, based on a sampling time of several minutes. 

\subsubsection{Electric Power Network (EPN)}

This is modeled using the following ac power flow equations, since the CEPDHN includes a distribution grid:
\begin{align}
	&P_{i,k}=\sum_{j=1}^{I}V_{i,k}V_{j,k}[B_{ij}\sin(\delta_{i,k}-\delta_{j,k})+G_{ij}\cos(\delta_{i,k}-\delta_{j,k})]
	\label{eq:PF1} \\
	&Q_{i,k}=\sum_{j=1}^{I}V_{i,k}V_{j,k}[G_{ij}\sin(\delta_{i,k}-\delta_{j,k})-B_{ij}\cos(\delta_{i,k}-\delta_{j,k})]
	\label{eq:PF2}
\end{align}
where $P_{i,k}$ and $Q_{i,k}$ denote the active and reactive power injected into the system at bus $i$ at time $k$ by all network participants $n$ connected at this bus. The system is assumed balanced for simplicity and without loss of generality \cite{alrumayh2019flexibility}. The following operational constraints are included for secure operation of the system:
\begin{IEEEeqnarray}{rCcCl}
	\underline{V}_{i} &\le& V_{i,k} &\le& \overline{V}_{i} \label{eq:eps1} \\
	\underline{P}_{n,k} &\le& P_{n,k} &\le& \overline{P}_{n,k} \label{eq:Pn}\\
	\underline{Q}_{n,k} &\le& Q_{n,k} &\le& \overline{Q}_{n,k}  \label{eq:eps8}
	\label{eq:Qn}
\end{IEEEeqnarray}
The limits on active and reactive power are defined by relevant distribution transformer capacities in the EPN and the EMSSAs of the FNPs, and may vary over time. Note that feeder limits can also be readily included in the proposed EPN model. 
\subsubsection{Energy Converters}

Energy converters such as heat pumps, electric boilers or CHP units are modeled by their respective coupling factors $\zeta$, as follows:
\begin{equation}
	\Phi_{m,k}=\zeta_{n} P_{n,k} \label{eq:ec}
\end{equation}

\subsubsection{District Heating Network (DHN)} \label{sec_DHN_model}
In the proposed model, the DPRs and pumps enable heat loss savings as the mass flows are optimally directed through the network \cite{LiSvendsenGudmundssonEtAl2017}, avoiding supplying consumers that are closest to heat sources with considerably larger differential pressures and mass flows than consumers further away. This prevents consumers close to producers to pass high water temperatures to the return network, and therefore reduces the heat losses of the DHN. By merging this model information with predictions on power infeed and demand and adequate controls, it is possible to fully exploit the flexibilities of the DHN, resulting from DSM and pipeline storage, for the EPN. 
\paragraph{Hydraulic Model}
The hydraulic model used in this work is described in detail in \cite{Oppelt2015,OppeltUrbaneckGrossEtAl2016,Koecher2000}, and defines the mass flows and differential pressures throughout the DHN. It comprises the continuity of flow which assures that in a closed system the amount of mass flow entering a node will also leave it, i.e. : 
\begin{equation}
	\boldsymbol{A} \boldsymbol{\dot{m}}_k=\boldsymbol{0} \label{eq:hyd1}
\end{equation}
where the edge-node incidence matrix $\boldsymbol{A}$ is multiplied with the vector of all edge mass flows $\boldsymbol{\dot{m}}$ in the network. The loop pressure equation, with the edge-loop incidence matrix $\boldsymbol{B}$ and the vector of edge pressure differences $\Delta\boldsymbol{p}$, represents the sum of all pressure differences in a loop adding up to zero, as follows:
\begin{equation}
	\boldsymbol{B} \Delta\boldsymbol{p}_k=\boldsymbol{0}
\end{equation}

Pumps and DPRs, which allow reducing heat losses provide predefined pressure values $\Delta \boldsymbol{p}^{\mathrm{def}}$ over a certain control path, which is a sequence of edges where the pressure is measured before and after; These control paths can be found over consumer facilities, as their flow control valves can control the mass flow more precisely if the valve differential pressure changes slightly. The input of the controls of DPRs and pumps on the pressure differences of the respective edges in the networks can be taken into account as follows:
\begin{equation}
	\boldsymbol{R} \Delta\boldsymbol{p}_k=\Delta\boldsymbol{p}_k^{\mathrm{def}} \label{eq:hyd3},
\end{equation}
where $\boldsymbol{R}$ is the edge control path matrix and its elements are defined as follows:
\begin{equation}
	r_{e}^{\mathrm{cp}} = \begin{dcases}
		~1, &\text{if $\mathrm{e}$ and  $\mathrm{cp}$ same direction} \\
		-1, &\text{if $\mathrm{e}$ and $\mathrm{cp}$ opposite direction} \\
		~0,&\text{if $\mathrm{e}$ not in $\mathrm{cp}$}\\
	\end{dcases} 
\end{equation}

The relation of mass flow and differential pressure on the edges of the network is dependent on different components of the DHN, as shown in Table \ref{tab:druckdiff}, where the friction factors $\mu$ are assumed calculated in advance based on the Colebrook equation, and initial mass flows for simplicity reasons here. An exact calculation of these factors within the optimization problem is found, for example in \cite{Troester1999}. The pressure loss over a control valve is dependent on the flow coefficient $K_{\mathrm{v}}$, the reference differential pressure $\Delta p_0$, and the reference density $\rho_0$.

\renewcommand{\arraystretch}{1.8}
\begin{table}
	\begin{center}
		\caption{Differential pressure over different component types \cite{Oppelt2015,Koecher2000}.} \label{tabelle_druckdifferenzen}
		\label{tab:druckdiff}
		\begin{tabular}{|lcc|}
			\hline
			Type of component & \hspace{0.5cm} & $\Delta\boldsymbol{p}_k = \boldsymbol{\varphi}(\boldsymbol{\dot{m}}_k)$ \\
			\hline
			\hline
			Controlled pump & & $\Delta p_{k}^{\mathrm{pump}}$ \\
			Pipeline & & $\mu_{\mathrm{l}}\dot{m}^2_k$\\
			Producer & & $\mu_{\mathrm{p}} \dot{m}^2_k$\\
			Control valve & & $\frac{\Delta p_0}{K_{\mathrm{v,k}}^2 \rho_0 \rho} \dot{m}^2_k$\\
			Differential pressure regulator & & $\Delta p_{k}^{\mathrm{dpr}}$ \\
			Consumer & & $\mu_{\mathrm{c}} \dot{m}^2_k$\\
			\hline
		\end{tabular}
	\end{center}
\end{table}
\renewcommand{\arraystretch}{1}

\paragraph{Thermal Model}

This describes the temperature changes in the different network components. Thus, assuming perfect mixing within a node, the temperature at node $T_i$ can be determined by calculating the weighted sum of all temperatures $T_{e_i}^{\mathrm{out}}$ of the mass flows coming out of the edges $e_i$ at node $i$, as follows:
\begin{equation}
	\sum_{e_i=1}^{I_{\mathrm{e},i}^{\mathrm{out}}}\dot{m}_{e_i,k}T_{i,k} = \sum_{e_i=I_{\mathrm{e},i}^{\mathrm{out}}+1}^{I_{\mathrm{e},i}^{\mathrm{out}}+I_{\mathrm{e},i}^{\mathrm{in}}}\dot{m}_{e_i,k}T_{e_i,k}^{\mathrm{out}} 
	\label{eq_opt_knoten_waermenetz}
\end{equation}
The cardinality of the respective edges leaving and entering this node are $I_{e,i}^{\mathrm{out}}$ and $I_{e,i}^{\mathrm{in}}$. The pipeline model used here is an adapted version of the well-known node method \cite{Benonysson1991}; improvements are described in detail next with the corresponding equations. It is important to mention that \cite{Benonysson1991} contains an equation which is used to model the impact of the steel core of the pipelines; as future DHNs will be operated at lower temperatures, steel cores will not be necessary, and thus these are neglected here. 

The lossless outlet temperature $T_{l}^{\mathrm{out_1}}$ of a pipeline $l$ is calculated by a weighted sum of temperatures $T_i$ of the water masses that have entered the pipeline several time steps before, as follows:
\begin{align}
	T_{l,k}^{\mathrm{out_1}} &=& \frac{1}{\dot{m}_{l,k} \Delta k} \bigg[\Big(R_{l,k} - \rho A_l L_l\Big) T_{i,k - \gamma_{l,k}} 
	+ \sum_{\nu = k - \varepsilon_{l,k} + 1}^{k - \gamma_k - 1} \Big(\dot{m}_{l,\nu} \Delta k T_{i,\nu}\Big)  \label{eq_t_out1} \nonumber \\
	&&+ \Big(\dot{m}_{l,k} \Delta k + \rho A_l L_l - S_{l,k}\Big) T_{i,k - \varepsilon_{l,k}} \bigg] 
\end{align}
\begin{figure}
	\centering
	\includegraphics[width=0.8\textwidth]{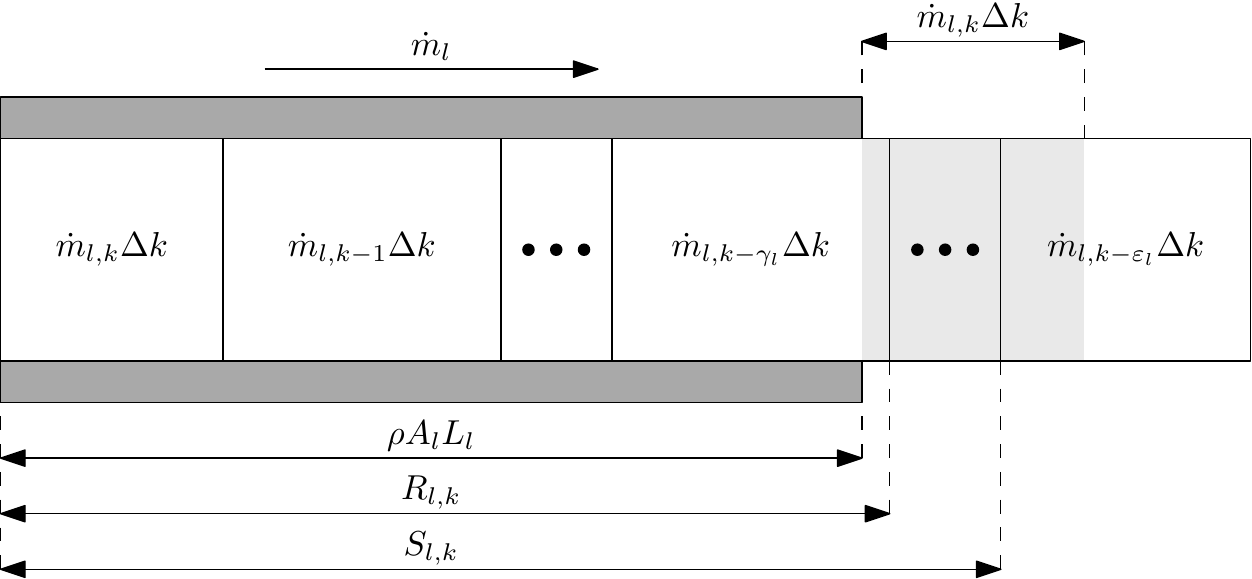}
	\caption{Scheme of a pipeline and relevant variables in the node method \cite{Benonysson1991}.}
	\label{fig_node_method}
\end{figure}
The water mass in the pipeline is the product of the density $\rho$, cross section area $A$, and length $L$ of the pipeline. The time step $\Delta k$ is then used with the mass flow $\dot{m}_{l,k}$ to define the total water mass leaving the pipeline at a time step. The three parts of the weighted mass (see Figure  \ref{fig_node_method}) are described by the following auxiliary variables, determining water masses $R$ and $S$ and time steps $\varepsilon$ and $\gamma$, that are relevant to determine these water masses: 
\begin{align}
		\gamma_{l,k} = \underset{x}{\mathrm{min}} \Bigg\{ x \text{ w.r.t. } &\sum_{\nu = 0}^{x} \left( \dot{m}_{l,k-\nu}\Delta k\right) \geq \rho A_l L_l, \ x \geq 0,\ x\in \mathbb{Z} \Bigg\} \label{eq_gamma_knotenmethode} \\
		\varepsilon_{l,k} = \underset{z}{\mathrm{min}} \Bigg\{ z \text{ w.r.t. } &\sum_{\nu = 1}^{z} \left( \dot{m}_{l,k-\nu}\Delta k\right) \geq \rho A_l L_l,\ z \geq 0,\ z\in \mathbb{Z} \Bigg\} \label{eq_epsilon_knotenmethdoe}
\end{align}
With the results of $\gamma_l$ and $\varepsilon_l$, the water masses $R_l$ and $S_l$ can then be calculated using:
\begin{align}
		R_{l,k} &= ~\sum_{\nu = 0}^{\gamma_{l,k}} \left( \dot{m}_{l,k-\nu}\Delta k \right) \label{eq_R}\\
		S_{l,k} &= \begin{cases} \sum_{\nu = 0}^{\varepsilon_{l,k} - 1} \left( \dot{m}_{l,k-\nu}\Delta k\right), &\text{ if } \varepsilon_{l,k} \geq \gamma_{l,k} + 1\\
			R_{l,k}, &\text{ else}\end{cases} \IEEEeqnarraynumspace
		\label{eq_S}
\end{align}

The final lossy outlet temperature of a pipeline $T_{l}^{\mathrm{out_2}}$ is, among others, dependent on the lossless output temperature $T_{l}^{\mathrm{out_1}}$, the length of stay $t_l$ of the water masses in the pipeline, the ambient temperature $T^\mathrm{a}$ and the thermal resistance $R'$ of the pipeline, as follows:
\begin{align}
	T_{l,k}^{\mathrm{out_2}} = T^\mathrm{a}_k 
	+ \Big( T_{l,k}^{\mathrm{out_1}} - T^\mathrm{a}_k \Big) \exp \Bigg( -\frac{1}{ \rho c_{\mathrm{p}} A_l R'_l } t_{l,k} \Bigg)  
	\label{eq_opt_temp_rohr3}
\end{align}

The value of $t_l$ is calculated using the following weighted sum, similar to (\ref{eq_t_out1}) \cite{MaurerElsnerKrebsEtAl2018}:
\begin{align}
	t_{l,k} =& \frac{1}{\dot{m}_{l,k} \Delta k} \bigg[ \gamma_{l,k} \Big( R_{l,k} - \rho A_l L_l \Big)  +~\sum_{\nu = k - \varepsilon_{l,k} + 1}^{k - \gamma_k - 1} (k - \nu) \dot{m}_{l,\nu} \Delta k  \nonumber \\
	&+ \varepsilon_{l,k} \Big( \dot{m}_{l,k} \Delta k + \rho A_l L_l - S_{l,k} \Big) \bigg] \Delta k
	\label{eq_verweildauer}
\end{align}

The auxiliary variables $\gamma_l$ and $\varepsilon_l$ turn the optimization problem into an MINLP problem, which is computationally expensive. Therefore, auxiliary weighting variables $\omega$ are introduced, to transform the model into an NLP problem, which are recalculated for the new mass flow values, over the prediction horizon, after every optimization in the ISOEMS: 
\begin{subequations}
	\begin{IEEEeqnarray}{rCl}
		\omega_{l,1,k} &=& \frac{1}{\dot{m}_{l,k}^\mathrm{pre} \Delta k} (R_{l,k} - \rho A_l L_l) \label{eq_param_approx_1} \\
		\omega_{l,2,k} &=& \frac{1}{\dot{m}_{l,k}^\mathrm{pre} \Delta k} (S_{l,k} - R_{l,k} ) \label{eq_param_approx_2} \\
		\omega_{l,3,k} &=& \frac{1}{\dot{m}_{l,k}^\mathrm{pre} \Delta k} (\dot{m}_{l,k}^\mathrm{pre} \Delta k + \rho A_l L_l - S_{l,k}) \IEEEeqnarraynumspace
		\label{eq_param_approx_3}
	\end{IEEEeqnarray}
\end{subequations}
These values are then inserted in (\ref{eq_t_out1}) and (\ref{eq_verweildauer}), with the approximation yielding adequate results, as these are based on the information from the previous time step. The authors have observed this in a test system, where the results of the full MINLP problem are similar to the ones obtained with the approximated NLP model \cite{MaurerRaRatzelMalanEtAl2021}.

The output temperature of a producer $T_{n}^\mathrm{out}$ is dependent on the thermal power $\Phi$ inserted by the producer, the temperature of the water flowing into the producer $T_{i}^\mathrm{r}$, and the respective mass flow $\dot{m}_n$, and can be represented as follows:
\begin{equation}
	T_{n,k}^\mathrm{out}= T_{i,k}^\mathrm{r} + \frac{\Phi_{n,k}}{c_{\mathrm{p}}\dot{m}_{n,k}} 
	\label{eq:prod}
\end{equation}
On the other hand, the temperature of water masses leaving a consumer is defined by characteristic curves, which are dependent on the temperature of the inflowing water masses $T_{i}^\mathrm{s}$, the ambient temperature $T^{\mathrm{a}}$, and the thermal power consumption of the consumers $\Phi_n$, and can be defined as follows: 
\begin{equation}
	T_{n,k}^\mathrm{out} = f\Big(T_{i,k}^\mathrm{s},T^{\mathrm{a}}_k,\Phi_{n,k}\Big) 
\end{equation}
Furthermore, the mass flow through the heat exchanger of the consumer $\dot{m}_{n,k}$ can be obtained from:
\begin{equation}
	\dot{m}_{n,k} = \frac{\Phi_{n,k}}{c_{\mathrm{p}} (T_{i,k}^\mathrm{s} - T_{n,k}^\mathrm{out})} \label{eq:massflowCons}
\end{equation}
Finally, the thermal impact of all components not explicitly stated before, are assumed to be negligible, and can therefore be modeled by:
\begin{equation}
	T_{c,k}^\mathrm{out} = T_{i,k} 
	\label{eq:comp}
\end{equation}

Operational constraints for a safe and reliable operation of the DHN are also required. Hence, similar to (\ref{eq:Pn}) and (\ref{eq:Qn}), the power infeed and demand of DHN participants are constrained by:
\begin{equation}
	\underline{\Phi}_{n,k} \leq \Phi_{n,k}\leq \overline{\Phi}_{n,k} 
	\label{eq_opt_unb_phi}
\end{equation}
Additionally, the possible change of output temperature of DHN participants feeding heat into the DHN are limited to prevent material fatigue as follows:
\begin{IEEEeqnarray}{rCl}
	T_{n,k-1}^\mathrm{out}- \Delta T^{\mathrm{prod}} \leq T_{n,k}^\mathrm{out}  \leq T_{n,k-1}^\mathrm{out}+ \Delta T^{\mathrm{prod}} \label{eq:ramp_constr}
\end{IEEEeqnarray}
All mass flows $\dot{m}$ need to be within defined bounds, with the lower bounds preventing the accumulation of deposits, and the upper bounds preventing high pressure losses and thereby lower the power necessary for operation of the pumps; thus:
\begin{equation}
	\underline{\dot{m}}_{e} \leq \dot{m}_{e,k} \leq \overline{\dot{m}}_{e}
	\label{eq_opt_unb_massflow1} 
\end{equation}
The lower bounds $\underline{\dot{m}}_{e}$ also enable to model the effects of check valves, which prevent a flow reversal at different components in the DHN as, for example, within consumer or producer facilities. The differential pressures $\Delta p$ over all edges need to be maintained within respective bounds, since the lower limits prevent evaporation and corrosion and guarantee satisfactory mass flow to meet the power demand of consumers, and upper limits prevent damage to network components; therefore:
\begin{equation}
	\underline{\Delta p}_{\mathrm{e}} \leq \Delta p_{e,k} \leq \overline{\Delta p}_{\mathrm{e}}
	\label{eq_opt_unb_maxdruck}
\end{equation}
Limits on valve set points are imposed by limiting the flow coefficient $K_{\mathrm{v}}$ as follows: 
\begin{equation}
	0 \leq K_{\mathrm{v}, n,k} \leq K_{\mathrm{vs}, n}. 
	\label{eq_opt_unb_maxkv} 
\end{equation}
Finally, the node temperatures $T_i$ are limited to guarantee satisfactory power provision to consumers and prevent damage of network components; thus:
\begin{equation}
	\underline{T}_{i} \leq T_{i,k} \leq \overline{T_{i}}
	\label{eq:dhn6}
\end{equation}

\subsection{Optimization Problem Implementation}

The resulting optimization problem can be written in the following form:
\begin{IEEEeqnarray}{rcl}
	& \underset{\boldsymbol{x}}{\mathrm{max}}\  &W(\boldsymbol{x}) \nonumber \\
	&\text{s.t.~~} &\boldsymbol{c}(\boldsymbol{x}) = \boldsymbol{0}  \\
	&& \boldsymbol{h}(\boldsymbol{x}) \geq \boldsymbol{0} \nonumber
	\label{eq_opt}
\end{IEEEeqnarray}
The objective function, which corresponds to \eqref{eq:SW}, is maximized subject to the equality constraints $\boldsymbol{c}(\boldsymbol{x})$, which correspond to the EPN equations \eqref{eq:PF1}, \eqref{eq:PF2}, the energy converters equation \eqref{eq:ec}, and the hydraulic and thermal model of the DHN represented by \eqref{eq:hyd1}-\eqref{eq:hyd3}, \eqref{eq_opt_knoten_waermenetz}, \eqref{eq_opt_temp_rohr3} and \eqref{eq:prod}-\eqref{eq:comp}. The inequality constraints $\boldsymbol{h}(\boldsymbol{x})$ consist of the EPN constraints \eqref{eq:eps1}-\eqref{eq:eps8}, and the DHN constraints \eqref{eq_opt_unb_phi}-\eqref{eq:dhn6}. It is important to mention that $W(\boldsymbol{x})$ incorporates all time steps of the prediction/rolling horizon.

The auxiliary variables in \eqref{eq_gamma_knotenmethode}-\eqref{eq_epsilon_knotenmethdoe} and \eqref{eq_param_approx_1}-\eqref{eq_param_approx_3}, used to solve this problem as an NLP problem, are precalculated with the predicted mass flows obtained from the previous time step optimization. Due to the receding horizon approach, the mass flows show considerable changes only when the behavior of network participants significantly changes from one time step to the next, which mainly takes place due to system contingencies. Therefore, the mass flows for the time step at the end of the prediction horizon are assumed to be identical to the previous time step. For a prediction horizon, the NMPC/ISOEMS should have a sufficient number of time steps to find proper values for the pre-calculated auxiliary variables, in the context of the NMPC implementation. 
Note, that consumer flexibility is not restricted by the precalculated mass flows, as these are only used within the thermal model, since in the hydraulic model, the mass flows are considered as optimization variables, and thus the flexibility services are fully available. The main impact of the pre-calculation of the mass flows in the proposed approach is that the node temperatures are approximated during the optimization process.

In addition to the bids and offers of producers and consumers, the differential pressures over the pumps are assumed to be adjusted by the operator and are submitted to the ISOEMS. This assumption reflects the present operating practice, which is the most challenging form of operation for the ISOEMS. A simpler operating approach would consist of giving the differential pressures of the pumps as set point ranges to the ISOEMS, which could then use these as further degrees of freedom within the optimization, respecting the given operational limits. Node temperatures and mass flows from previous time steps are obtained from measurements in the network and previous optimization time steps.

\section{Case Studies and Results} \label{case_study}
The performance of the proposed TC approach is studied considering two scenarios, which represent diverse functionalities of the presented TCS. The first scenario highlights how flexibility is included in CEPDHN operation by utilizing a BESS, demand side management of consumers in both networks, and the heat storage capabilities of the DHN. The second scenario examines how the operation of a CEPDHN changes for a very cold winter day with high infeed of renewable energy, thereby demonstrating how the coupled operation improves overall system operation in contrast to independent operation of the networks by a TCS, as reported in \cite{LiLianJ.ConejoEtAl2020}. Possible price signals for the EPN and the DHN are also discussed, as well as the impact of varying bid prices. 

\subsection{Scenario 1: DHN Flexibility Provision for the EPN}

\subsubsection{Electric Power Network}
\begin{figure}
	\centering
	\includegraphics[width=0.8\textwidth]{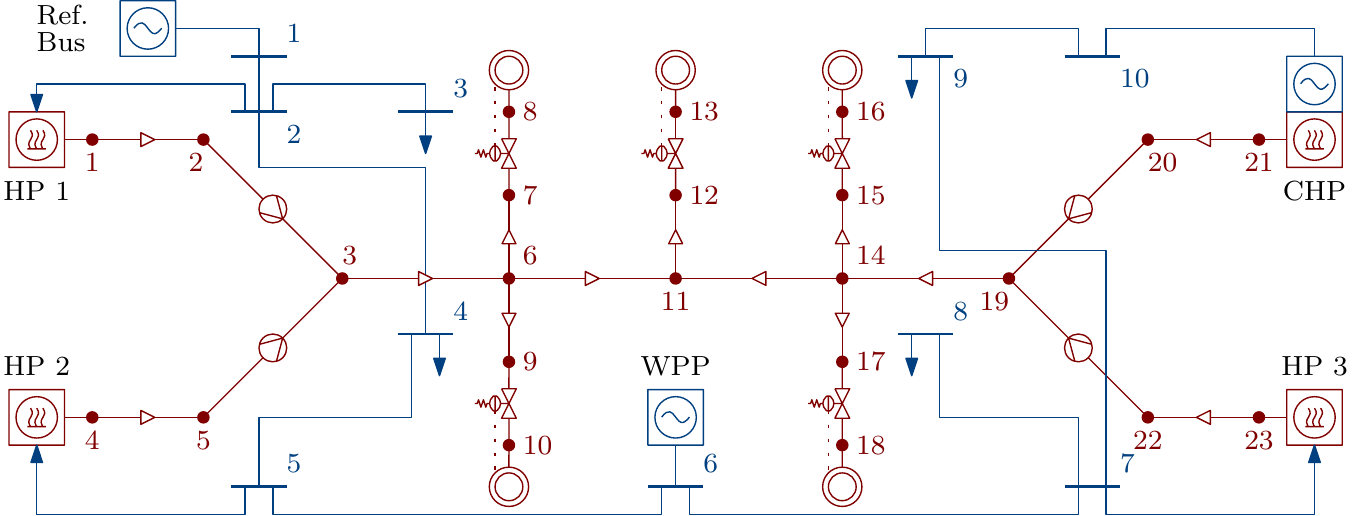} 
	\caption{Schematic diagram of the EPN (shown in blue) and the supply network of the DHN (drawn in red color); the symmetric DHN return network shown in Figure \ref{fig_dhn_network} is omitted here for simplicity.}
	\label{fig_network}
\end{figure}
The \SI{11}{\kilo\volt} distribution network from the case study in \cite{liu_diss14} is used here as part of the test system, and is shown in Figure  \ref{fig_network}. The EPN comprises ten buses and nine radial feeder sections. The flexible consumers, including the heat pumps, are located at Buses 2, 3, 4, 5, 7, 8, and 9. The producers and energy converters are chosen to clearly illustrate the energy flows in a future energy network.
The CHP represents a flexible producer whereas, the Wind Power Plant (WPP) is modeled as an inflexible producer, which aligns with the German regulation for maximum infeed of RES.
The distribution grid is assumed to have no power exchange with the transmission system; this is facilitated by placing a BESS at the reference bus, which can be considered as a challenging EPN operating condition. The heat pumps are considered as flexible consumers in the EPN and flexible producers in the DHN.

\subsubsection{District Heating Network}
\begin{figure}
	\centering
	\includegraphics[width=0.8\textwidth]{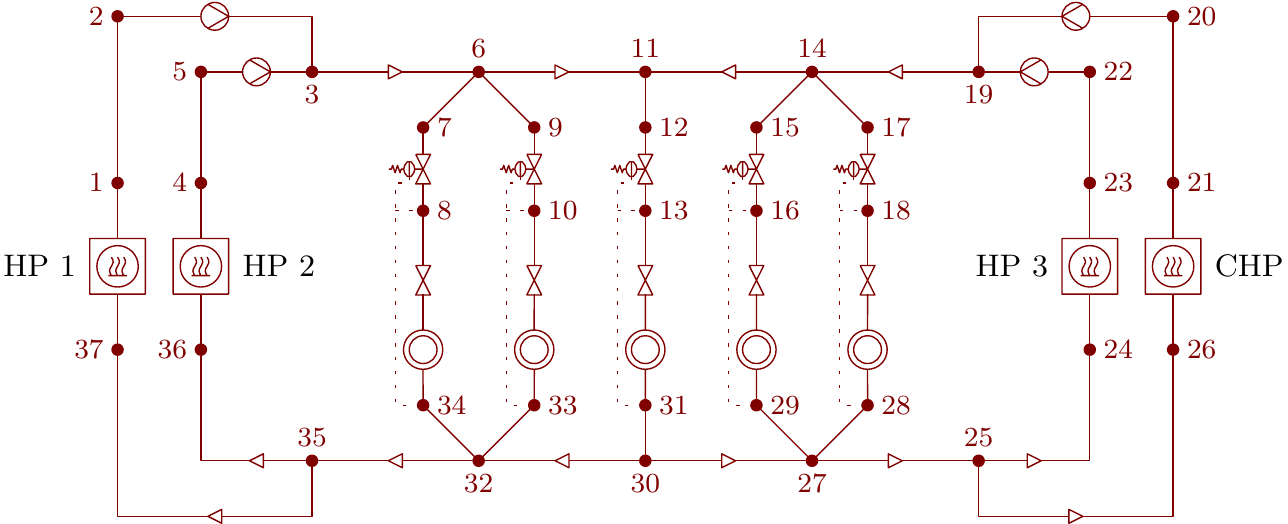}
	\caption{Schematic diagram of the meshed district heating network with supply and return network.}
	\label{fig_dhn_network}
\end{figure}

The DHN supplies all consumers with heat, injected by the HPs and the CHP, as shown in Figure \ref{fig_dhn_network}.
The FNP at Node 13 is a large-scale consumer (e.g. a secondary network) while the other FNPs are small-scale consumers. This DHN has a meshed structure due to the presence of the supply and return network; the flow directions are determined in advance through the operational states of the pumps and the set point ranges of the valves. The effects of check valves in consumer and producer facilities prevent flow reversals, which are taken into account in \eqref{eq_opt_unb_massflow1}. The pumps are located next to the producer facilities, i.e., between Nodes 2 and 3, and the DPRs are found in the consumer lines, such as the one between Nodes 7 and 8, which ensures that the pressure difference between Nodes 8 and 34 remains constant by varying the differential pressure between Nodes 7 and 8. The temperature limits which ensure safe and reliable operation of the supply network are set to \SI{80}{\degreeCelsius} and  \SI{130}{\degreeCelsius}, as per \cite{EnergieSchweiz2018}, and the minimum output temperature of all producers is assumed to be \SI{95}{\degreeCelsius}, for efficiency reasons. An ambient temperature of \SI{10}{\degreeCelsius} is considered, as this is a typical autumn temperature in Germany. The hydraulic network parameters, such as the roughness, diameters, and lengths of the pipelines, are taken from \cite{liu_diss14}.

\subsubsection{Grid Participants}
For simplicity and without loss of generality, inflexible consumers are not considered in both studies presented here. The system base power profile and associated peak demands are defined by assuming that consumers in the power system and DHN are not participating in the TCS auction market. It is further assumed that consumers are capable of reducing their power consumption up to $\SI{30}{\percent}$ at any point in time, which allows to properly demonstrate the performance of the proposed framework and the models; in practice, this flexibility would be determined by their EMSSAs by first calculating the predicted minimum power demand for the next hours and the expected usable amount of energy. The power profiles of consumers are based on the electric load profiles of \cite{ed_netze}, and the heat load profiles of \cite{yao_steemers}, scaling, time-shifting, and overlaying them with a noise profile to represent different consumers. The power profile of the WPP was taken from \cite{smard}, and was scaled down and shifted. The constant coupling factor of the CHP was set to \SI{1.42}{}, and the coupling factors of the heat pumps were set to 4, based on \cite{EnergieSchweiz2018}.

The cost and benefit parameters $c_{i}$ of producers and consumers are given in Table \ref{table_cost_ben_param}. The same values for $c_{i}$ are used here for all time steps $k$ in order to facilitate the analysis of the simulation outcome; however, results with varying cost parameters are also tested and discussed and discussed in Section \ref{sec:Res_Var_Bids}. As, only the bids and offers are sent to the ISOEMS by the FNPs, privacy is guaranteed here by design. It is important to mention that the power injection as well as power consumption of the reference bus/BESS results in income for the battery in the form of an ancillary service, which can be interpreted as an incentive for storage systems. \textcolor{black}{Assuming losses of $\SI{3}{\percent}$ for charging and discharging each in a Lithium-ion battery system, the battery used within the case study represents a storage system with at least 300~kWh capacity, 200~kWh initial charging and 120~kW maximum charging/discharging rate. However, as the battery represents an FNP, these system parameters are only known by this FNP. For example if losses increase do to the aging of the battery, the FNP can vary the offer prices in order to keep the revenue constant. }

\begin{table}
	\caption{Cost and benefit parameters of flexible grid participants in general monetary \\units \UseTextSymbol{TS1}{¤} for Scenario 1.}
	\label{table_cost_ben_param}
	\centering
	\begin{tabular}{|c|c||c|c|}
		\hline
		& & $c_{i}$ in \UseTextSymbol{TS1}{¤}/MW \\
		\hline
		\hline
		\multirow{4}{*}{\begin{turn}{90}\rlap{\hspace{-7pt}EPN}\end{turn}} & BESS & 25 \\
		& CHP & 30 \\
		& Heat pump & 12 \\
		& Flexible Consumer & 9 \\
		\hline
		\multirow{3}{*}{\begin{turn}{90}\rlap{\hspace{-7pt}DHN}\end{turn}} & CHP & 20 \\
		& Heat pump & 6 \\
		& Flexible Consumer & 7 \\
		\hline
	\end{tabular}
	\renewcommand{\arraystretch}{1}
\end{table}

\subsubsection{Results and Analysis}
The prediction horizon of the ISOEMS is set to 16 time steps, of \SI{15}{\minute} each, which is the interval used in the German intraday market, and the total simulation horizon is \SI{24}{\hour}. Several options can be considered using the proposed approach to provide additional flexibility to the power system. In particular and as demonstrated next, one option is to use DSM of electric and heat consumers to bridge a period with high power price. This leads to a cost advantage for the consumers, allowing to minimize the power injected by more expensive nonrenewable energy sources. Another option is the use of heat pumps during a period of high penetration with RESs to transfer energy to the DHN.

\begin{figure}
	\centering
	\footnotesize
	\input{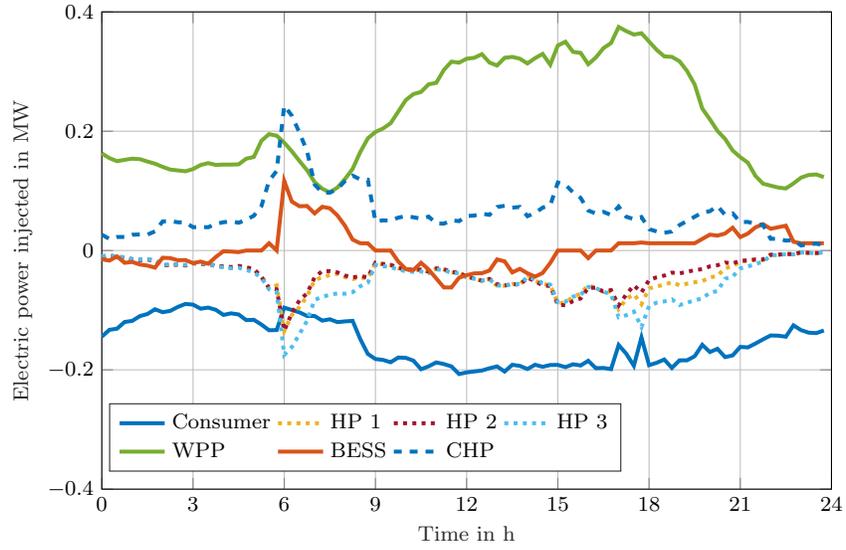}
	\caption{Injected power by all EPN participants for Scenario 1.}
	\label{fig_elec_power}
\end{figure}

\begin{figure}
	\centering
	\footnotesize
	\input{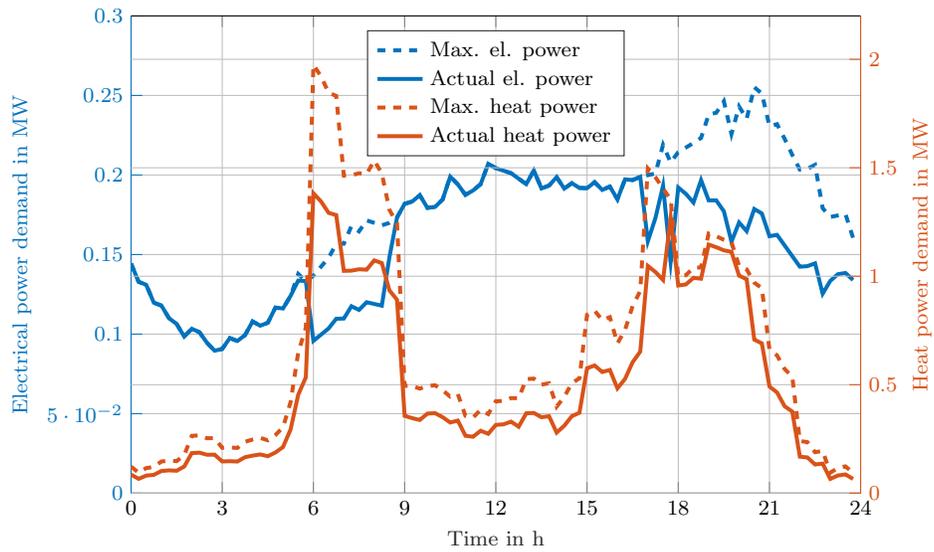}
	\caption{Aggregated electric and heat power demand of all consumers for Scenario 1.}
	\label{fig_dsm}
\end{figure}

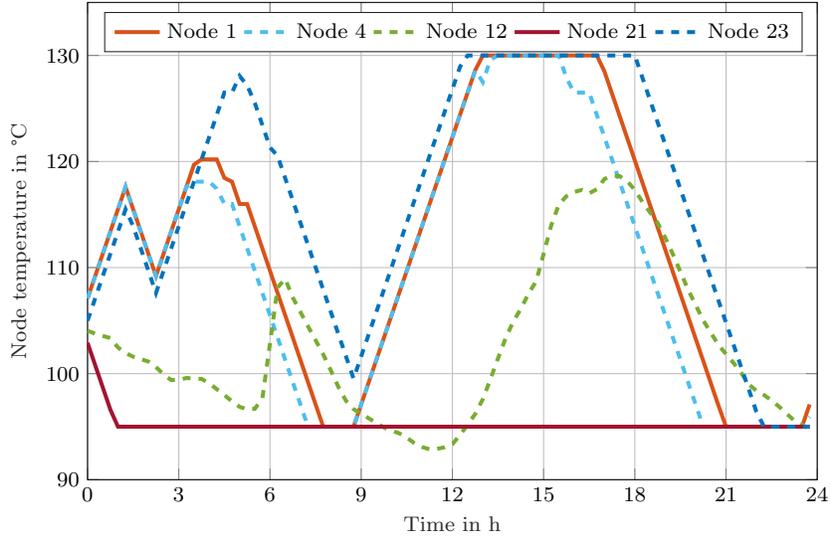
\begin{figure}
	\centering
	\footnotesize
%
%
\definecolor{mycolor1}{rgb}{0.85000,0.32500,0.09800}%
\definecolor{mycolor2}{rgb}{0.30100,0.74500,0.93300}%
\definecolor{mycolor3}{rgb}{0.46600,0.67400,0.18800}%
\definecolor{mycolor4}{rgb}{0.63500,0.07800,0.18400}%
\definecolor{mycolor5}{rgb}{0.00000,0.44700,0.74100}%
\begin{tikzpicture}

\begin{axis}[%
at={(0.758in,0.481in)},
scale only axis,
xmin=0,
xmax=24,
xtick={ 0,  3,  6,  9, 12, 15, 18, 21, 24},
xlabel style={font=\color{white!15!black}},
xlabel={Time in h},
ymin=90,
ymax=135,
ylabel style={font=\color{white!15!black}},
ylabel={Node temperature in \degree C},
axis background/.style={fill=white},
xmajorgrids,
ymajorgrids,
legend columns = 5,
legend style={legend cell align=left, align=left, draw=white!15!black, font=\footnotesize}
]
\addplot [color=mycolor1, line width=\mylinewidth]
  table[row sep=crcr]{%
0	107.099999886613\\
1.25	117.599985324629\\
2.25	109.199985522651\\
3.5	119.699984646077\\
3.75	120.19999999111\\
4.25	120.199999990823\\
4.5	118.44748028393\\
4.75	118.099999591831\\
5	115.999999995446\\
5.25	115.985309712385\\
6.25	107.59999997886\\
7.75	94.9999999907814\\
8.75	94.9999999911915\\
12.75	128.60000016526\\
13	129.999999869425\\
16.75	130.000000012721\\
17	128.501784280146\\
20.75	97.0017841554938\\
21	94.9999999908021\\
23.5	94.9999999956318\\
23.75	97.0999985456768\\
};
\addlegendentry{Node 1}

\addplot [color=mycolor2, dashed, line width=\mylinewidth]
  table[row sep=crcr]{%
0	107.099999874202\\
1.25	117.599984951796\\
2.25	109.19998512821\\
3.25	117.599979230059\\
3.5	118.099999990445\\
4	118.099999990082\\
4.25	117.432505365789\\
4.5	115.99999981242\\
4.75	115.99999946585\\
7.25	95.0000001402302\\
8.75	94.9999999911987\\
12.75	128.600000165216\\
13	127.496884815071\\
13.25	129.596884826291\\
13.5	130.000000012679\\
15.5	129.999999969297\\
15.75	127.899999958143\\
16	126.499999997425\\
16.5	126.499999991013\\
20.25	94.999999990652\\
23.5	94.9999999956246\\
23.75	95.93783180842\\
};
\addlegendentry{Node 4}

\addplot [color=mycolor3, dashed, line width=\mylinewidth]
  table[row sep=crcr]{%
0	104.046570555575\\
0.25	103.764672509727\\
0.5	103.551193331071\\
0.75	103.356476721225\\
1	102.466207865335\\
1.25	101.941907000221\\
1.5	101.596779892855\\
2	101.140410503122\\
2.25	100.625311923916\\
2.5	99.928760922961\\
2.75	99.3966620927685\\
3	99.4065651314253\\
3.25	99.5898346025348\\
3.5	99.5419889676175\\
3.75	99.5030356395263\\
4	99.0901080778768\\
4.5	97.9857702947809\\
4.75	97.3848918874321\\
5	96.8729583136326\\
5.25	96.6323963001891\\
5.5	96.6842387043652\\
5.75	97.6853972072877\\
6	102.832710364689\\
6.25	108.129224021912\\
6.5	108.934339392956\\
6.75	107.170147902245\\
7.25	104.457751212009\\
7.5	103.155830209687\\
7.75	101.811627279048\\
8	100.309777632301\\
8.25	98.7456286321721\\
8.5	97.4404317416761\\
8.75	96.6520544116924\\
9	96.2040061282924\\
9.25	95.7888004063524\\
9.5	95.3444497204003\\
9.75	94.9196990713944\\
10	94.6153609586035\\
10.25	94.3781433818091\\
10.5	93.9554625538979\\
10.75	93.4510109261584\\
11	93.0869570587841\\
11.25	92.8662280633865\\
11.5	92.910231704652\\
11.75	93.0601691310242\\
12	93.3418378839729\\
12.25	94.2300411807813\\
12.5	95.167326568469\\
12.75	96.2048358606149\\
13	97.5898713494246\\
13.25	99.4743564743317\\
13.5	101.446685487748\\
13.75	103.328802996971\\
14	104.905097773078\\
14.25	106.106673739697\\
14.5	107.531092932694\\
14.75	108.683024004058\\
15	111.279495751988\\
15.25	113.977304812875\\
15.5	115.933962514795\\
15.75	116.733321637536\\
16	117.12139990514\\
16.25	117.316607831781\\
16.5	117.003657781809\\
16.75	117.387249157314\\
17	118.351933460921\\
17.25	118.658106167812\\
17.5	118.626105494382\\
17.75	118.152851594167\\
18	117.317341028406\\
18.25	116.19427239104\\
18.75	114.212078519034\\
19	112.825515736297\\
19.25	111.269663327975\\
19.75	107.92602333122\\
20.25	105.021878660826\\
20.75	102.704578343394\\
21.25	100.99340325707\\
21.5	100.071537247724\\
21.75	98.9863788734937\\
22	98.3792287979087\\
22.25	97.9886952661034\\
22.5	97.4688171183161\\
22.75	96.8759812682378\\
23	96.2064886473351\\
23.25	95.5629327477675\\
23.5	95.1556402767649\\
23.75	94.7679316640233\\
};
\addlegendentry{Node 12}

\addplot [color=mycolor4, line width=\mylinewidth]
  table[row sep=crcr]{%
0	102.899999989771\\
0.75	96.6000000202698\\
1	94.9999999906952\\
23.75	94.9999999909288\\
};
\addlegendentry{Node 21}

\addplot [color=mycolor5, dashed, line width=\mylinewidth]
  table[row sep=crcr]{%
0	104.999366659082\\
1.25	115.499347307051\\
1.5	113.903756038308\\
2.25	107.603756103324\\
4.5	126.499999794113\\
4.75	126.499999625666\\
5	128.067945377312\\
5.25	127.171027185235\\
5.5	125.527423337079\\
6	121.327423312642\\
6.25	120.527144100711\\
8.75	99.5271440275311\\
12.25	128.927144184157\\
12.5	130.000000012615\\
18	130.00000001096\\
22	96.3999998626116\\
22.25	94.9999999912675\\
23.75	95.000000027594\\
};
\addlegendentry{Node 23}

\end{axis}
\end{tikzpicture}%
	\caption{Node temperatures in the DHN for Scenario 1.}
	\label{fig_node_temp}
\end{figure}

The electric power injected by EPN participants, calculated using the proposed approach, is shown in Figure \ref{fig_elec_power}, where the power injected by consumers is shown in aggregated form.
A more detailed view of the consumer power is illustrated in Figure \ref{fig_dsm} where the maximum and actual aggregated power of electrical and heat consumers is shown.
To understand the temporal progression of the electric consumer power, a comparison with the power of the reference bus and the WPP of Figure  \ref{fig_elec_power} is helpful. Thus, during a period with a large ratio between the infeed of the WPP and the aggregated demand, the power of the electrical consumers is maximized to minimize the power transferred by the reference bus; this is the case between \SI{0}{\hour} and \SI{5.25}{\hour}, and between \SI{9.00}{\hour} and \SI{16.75}{\hour}. The rest of the time, the consumed power is reduced to its assumed minimum of \SI{30}{\percent} of the maximum power demand. This allows minimizing the power injected by the reference bus and the CHP in the periods where the injection of the WPP cannot provide the entire power demand. 
To illustrate the behavior of the heat pumps, the output temperatures at Node 1, 4, and 23, in Figure \ref{fig_node_temp} can be used in addition to Figure  \ref{fig_elec_power}.
In contrast to the consumers, the power of a heat pump is coupled with the output temperature through the temperature ramping constraint \eqref{eq:ramp_constr}; this causes a strong correlation to the current mass flow flowing through the heat pump, as per \eqref{eq:prod}. Thus, during the period between
\SI{8.75}{\hour} and \SI{15.5}{\hour}, the heat pumps consume maximum power; a further increase of their thermal power output is not possible due to the temperature limits, as depicted in Figure \ref{fig_node_temp}. 
During the periods between \SI{7.75}{\hour} and \SI{8.75}{\hour} and \SI{22.25}{\hour} and 23.5\hspace{1.5pt}h, the power consumption of the heat pumps is minimal. The sharp changes in the temperature are due to the varying wind power infeed, as may be noted in Figure \ref{fig_elec_power}, and the varying electric and heat power demand, as per Figure \ref{fig_dsm}.

\begin{figure}
	\centering
	\footnotesize
%
%
\definecolor{mycolor1}{rgb}{0.46600,0.67400,0.18800}%
\definecolor{mycolor2}{rgb}{0.85000,0.32500,0.09800}%
\definecolor{mycolor3}{rgb}{0.00000,0.44700,0.74100}%
\begin{tikzpicture}

\begin{axis}[%
axis y line*=left,
at={(0.758in,0.481in)},
scale only axis,
xmin=0,
xmax=24,
xtick={ 0,  3,  6,  9, 12, 15, 18, 21, 24},
xlabel style={font=\color{white!15!black}},
xlabel={Time in h},
ymin=-0.4,
ymax=1,
ylabel style={font=\color{white!15!black}},
ylabel={Power injected in MW},
axis background/.style={fill=white},
xmajorgrids,
ymajorgrids,
]
\addplot [color=mycolor1, line width=\mylinewidth]
  table[row sep=crcr]{%
0	0.162899999900954\\
0.5	0.149790000512816\\
1.25	0.153209999900671\\
2.75	0.132879999900887\\
4	0.144089999911269\\
4.5	0.14447000054712\\
5	0.156812796006605\\
5.25	0.183987977407124\\
5.5	0.195148333432222\\
5.75	0.192488333323837\\
6.25	0.16633166674972\\
7.25	0.103600000099565\\
7.5	0.097393333432894\\
8	0.12000333343283\\
8.25	0.136850000099201\\
8.75	0.188055000098878\\
9.75	0.232819999900798\\
10	0.252199999900572\\
10.75	0.278609999900564\\
11	0.281079999900484\\
11.25	0.302739999900449\\
11.5	0.31660999990045\\
11.75	0.314899999900486\\
12.5	0.329149999900519\\
12.75	0.31508999990055\\
13	0.310149999910003\\
13.25	0.323069999900802\\
13.75	0.32154999995937\\
14	0.313569999900519\\
14.5	0.329529999900522\\
14.75	0.311479999900726\\
15	0.343730973046856\\
15.25	0.349995082870212\\
15.5	0.332904138797449\\
15.75	0.331355773620011\\
16	0.312239999901067\\
16.75	0.347762412882187\\
17	0.374698333432221\\
17.5	0.362063333431937\\
17.75	0.36428000003983\\
18.5	0.325488333392091\\
19	0.317286666656649\\
19.5	0.278273333147848\\
19.75	0.237930000099237\\
20.75	0.167661666764875\\
21.25	0.147268333432518\\
21.5	0.125101666762291\\
21.75	0.112023333432841\\
22.5	0.104043333432831\\
23.25	0.126653333432223\\
23.75	0.123106666765555\\
};
\label{plot_one}

\addplot [color=mycolor3, line width=\mylinewidth]
  table[row sep=crcr]{%
0	0.0736468258423457\\
0.25	0.0601989335136359\\
0.5	0.0779533279243942\\
0.75	0.0854586957707326\\
1	0.10897538015557\\
1.25	0.119736187775995\\
1.5	0.111463421693912\\
1.75	0.125268567885364\\
2	0.178329845927628\\
2.25	0.168618826724607\\
2.5	0.172966822430027\\
2.75	0.187188555092906\\
3	0.164067757270093\\
3.5	0.180363943134044\\
3.75	0.209795972873884\\
4.25	0.234791388965569\\
4.5	0.220473054578047\\
5	0.266335594795034\\
5.25	0.349090870647171\\
5.5	0.473656270268634\\
5.75	0.426602982389387\\
6	0.739547633247128\\
6.25	0.475467808751542\\
6.5	0.262157049534832\\
6.75	0.0201421954925216\\
7	-0.156548642814723\\
7.25	-0.26246806154705\\
7.5	-0.294089986031153\\
7.75	-0.293216497195044\\
8	-0.271173268018639\\
8.25	-0.231314560914758\\
8.5	-0.172840327027423\\
8.75	-0.141031149338243\\
9	-0.0246311879218553\\
9.25	0.0131075232889657\\
9.5	0.0387867753699993\\
10	0.107811303256163\\
10.75	0.183914211811995\\
11	0.169227918330051\\
11.25	0.186410545348224\\
11.5	0.22851233075707\\
11.75	0.242243850858294\\
12	0.302147708323112\\
12.25	0.329404502199086\\
12.5	0.36524899937983\\
12.75	0.35386509911331\\
13	0.431011322745722\\
13.25	0.444808307460509\\
13.5	0.424541849685681\\
13.75	0.431468222237893\\
14	0.339989582897751\\
14.5	0.423105728148585\\
14.75	0.434691803709008\\
15	0.650198112708427\\
15.25	0.627192881591238\\
15.5	0.538489851044144\\
15.75	0.464701285032415\\
16	0.338811086806334\\
16.25	0.314682722839269\\
16.5	0.317005006877235\\
16.75	0.266027352675536\\
17	0.265949236585389\\
17.25	0.114749303996984\\
17.5	0.0274289514871882\\
17.75	-0.0362506381148933\\
18	-0.0812429621689432\\
18.5	-0.211011663686548\\
18.75	-0.255812456383921\\
19	-0.333384670552181\\
19.5	-0.361307490620085\\
19.75	-0.370158628678617\\
20	-0.339858412637231\\
20.25	-0.338667485600059\\
20.5	-0.235328078941855\\
20.75	-0.232639832345019\\
21	-0.164275787716331\\
21.25	-0.15075984955724\\
21.5	-0.126508990302302\\
21.75	-0.114694599600931\\
22	-0.0508329602978392\\
22.25	-0.0494809592036027\\
22.5	-0.0377840749459715\\
22.75	-0.0352791172173923\\
23	-0.0159735447326597\\
23.5	-0.0190451906046043\\
23.75	-0.0113408249583742\\
};
\label{plot_two}

\end{axis}

\begin{axis}[
	axis y line*=right,
	axis x line=none,
	at={(0.758in,0.481in)},
	scale only axis,
	xmin=0,
	xmax=24,
	ymin=80,
	ymax=130,
	xtick={ 0,  3,  6,  9, 12, 15, 18, 21, 24},
	xlabel style={font=\color{white!15!black}},
	xlabel={Time in h},
	ylabel=Average node temperature in \degree C,
	ylabel style={font=\color{mycolor2}},
	every outer y axis line/.append style={mycolor2},
	every y tick label/.append style={font=\color{mycolor2}},
	every y tick/.append style={mycolor2},
	ymajorgrids,
	legend columns = 2,
	legend style={at={(0.5,0.97)}, anchor=north, legend cell align=left, align=left, draw=white!15!black}
	]
	\addlegendimage{/pgfplots/refstyle=plot_one}\addlegendentry{Power of WPP}
	\addlegendimage{/pgfplots/refstyle=plot_two}\addlegendentry{Heat power balance}

\addplot [color=mycolor2, dashed, line width=\mylinewidth]
table[row sep=crcr]{%
	0	91.0667643398127\\
	0.25	89.8796609746318\\
	0.5	89.1264234258591\\
	0.75	88.6192303436098\\
	1	88.7427520064857\\
	1.25	88.923528508094\\
	1.5	89.1591025494253\\
	1.75	89.3378240028394\\
	2	89.2363194294501\\
	2.25	89.1036257551937\\
	2.75	89.4242972832088\\
	3	89.5701376133206\\
	3.25	89.6700414065641\\
	3.5	89.5523865990457\\
	3.75	89.2673138390442\\
	4	89.0151557556325\\
	4.25	88.8591283007613\\
	4.5	89.0208843783225\\
	4.75	89.5222240116617\\
	5	90.0026487105601\\
	5.25	91.4544940794536\\
	5.5	92.1630489299972\\
	5.75	93.8768000957195\\
	6	97.1827919946517\\
	6.25	98.7330203687746\\
	6.5	98.1943596145641\\
	6.75	97.2783002016088\\
	7	96.2983700559431\\
	7.25	95.1937209280693\\
	7.5	93.9659749553012\\
	8	91.269108369636\\
	8.25	90.0401482589061\\
	8.5	88.965113766709\\
	8.75	88.1342326318898\\
	9	87.7424614595146\\
	9.25	87.4219879773299\\
	9.5	87.2150935496468\\
	9.75	87.2322972745261\\
	10.25	87.0775055273991\\
	10.5	87.0342057680901\\
	10.75	87.1560850867409\\
	11	87.4716416278158\\
	11.25	88.0479092193166\\
	11.5	88.6510913583402\\
	11.75	89.4080725123535\\
	12	90.1970428853064\\
	12.25	91.1914308094365\\
	12.5	92.4437975673243\\
	12.75	93.3692299651104\\
	13	94.4700659051655\\
	13.25	95.7298916148579\\
	13.5	96.8796014213103\\
	14	99.3285363110667\\
	14.25	100.656766996678\\
	14.5	102.089280609626\\
	14.75	103.091071112478\\
	15	104.273489583913\\
	15.25	105.55465429671\\
	15.5	106.49902803961\\
	15.75	107.277215508856\\
	16	107.857143112316\\
	16.25	108.455690684481\\
	16.5	108.882657219668\\
	16.75	109.185492604675\\
	17	109.561795570121\\
	17.25	109.660466816995\\
	17.5	109.384914374054\\
	17.75	108.834697997894\\
	18	108.012651172348\\
	18.25	107.113205449736\\
	18.5	106.12388775725\\
	18.75	104.964276027714\\
	19	103.565204830805\\
	19.75	98.7839485250944\\
	20	97.2723160019813\\
	20.25	95.8229418482393\\
	20.5	94.5173625231568\\
	21	92.3113344233927\\
	21.5	90.4525305176531\\
	21.75	89.3410373533583\\
	22	88.5568969393842\\
	22.25	87.8994332851522\\
	22.5	87.2057231579928\\
	22.75	86.6492233503\\
	23	85.7352057245547\\
	23.25	85.1054431541615\\
	23.5	84.6580537748044\\
	23.75	84.3570336133102\\
};
\addlegendentry{Average node temperature}

legend style={legend cell align=left, align=left, draw=white!15!black}

\end{axis}
\end{tikzpicture}%
	\caption{Power balance and average node temperature in the DHN and injected power of the WPP in the EPN for Scenario 1.}
	\label{fig_balpower_averagetemp}
\end{figure}
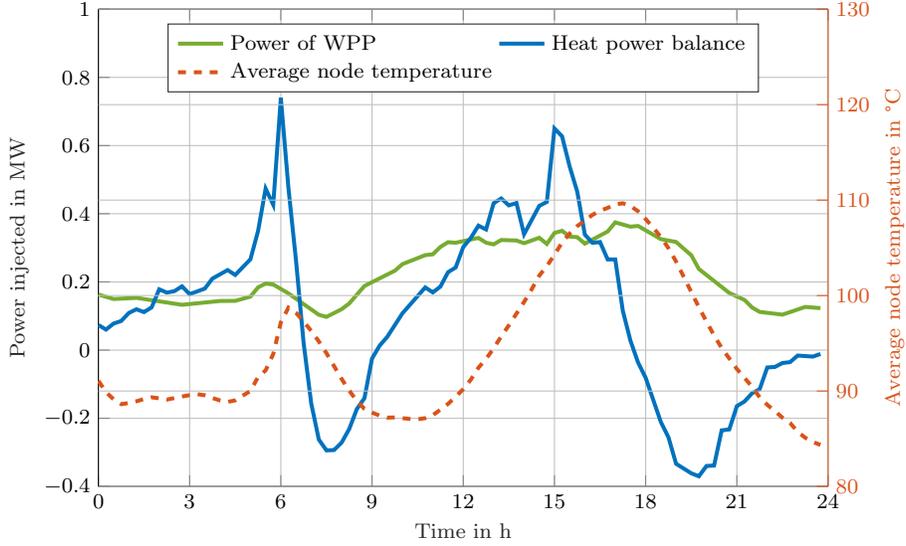

The effective flexibility provision by the DHN can be illustrated in Figure \ref{fig_balpower_averagetemp},
where the balance between the aggregated power injection and consumption in the DHN can be observed. The average node temperature and the power of the WPP is also shown to highlight the correlation to the heat power balance. This Figure illustrates the energy that can be stored in the DHN; thus, the DHN provides flexibility to the EPN because excess power from it can be transferred to the DHN, supplying heat consumers with heat power. 
In Figure \ref{fig_balpower_averagetemp}, it can also be observed, that the TCS achieves a very efficient form of DHN operation with low losses, as temperatures in the DHN are continually reduced, during periods when low cost power infeed of the WPP is sparely available. The described effect is observed from the profile of the WPP power infeed and the decreasing average node temperature, with a maximum decline of \SI{30}{\degreeCelsius}, between \SI{17.5}{\hour} and \SI{24}{\hour}. This is particularly relevant, as reducing the temperatures has the largest impact on reducing heat losses in a DHN \cite{ComakliYuekselComakli2004}. 

By design, the TCS inherently minimizes the losses in the CEPDHN, which results from two important aspects. The first is that the prediction horizon of the underlying NMPC control mechanism is limited; thereby, the ISOEMS is always incentivized to reduce the temperatures in the DHN to the operational limits to the end of the prediction horizon, in order to maximize the welfare in \eqref{eq:SW} by reducing the allocated power infeed into the DHN. Therefore, the TCS always directly uses the stored energy in the DHN pipelines as soon as possible, for welfare maximization. The second is the simple fact that lower network losses in the entire CEPDHN need less power infeed, which also inherently incentivizes the ISOEMS to reduce losses to maximize \eqref{eq:SW}. These aspects depict how market and control mechanisms of the proposed TC approach optimally collaborate to achieve technical efficiency and economic optimality. 

The optimization was implemented in GAMS using the IPOPT solver \cite{NocedalWright2006}. After the initialization, each simulated time step, which includes the 16 prediction horizon time steps, took less than $\SI{2}{\min}~ \SI{28.97}{\second}$ of CPU time. The entire simulation of all 96 time steps for this scenario took $\SI{76}{\min}~ \SI{35.20}{\second}$ on an Intel i7-6600U CPU with 2.60 GHz, which enables computation in real time even without exploiting the possibilities of distributed parallel optimization or faster workstations.

\subsection{Scenario 2: Price Signals, Independent versus Coupled Operation of EPN and DHN, and Varying Bid Prices}
\subsubsection{EPN, DHN, and FNPs}
The second scenario depicts a cold winter day with an ambient temperature of $T_{\mathrm{a}}$ of $\SI{-10}{\celsius}$. The BESS at the reference bus in Figure \ref{fig_network} is replaced by a Photovoltaic (PV) plant, and the profile of the WPP corresponds to data for the simulated day; both RES profiles where taken from \cite{smard} and scaled down to match the demand profiles. Furthermore, RESs are treated as all other flexible producers and can thus be curtailed, which results in all auction bids coming from fully flexible FNPs, that can reduce their demand and supply by $\SI{100}{\percent}$. Inflexible demand is assumed to be allocated in earlier cleared markets or long-term contracts. To show the operation for different hydraulic conditions, the operating points of pumps can be set by the TCS in this scenario. Finally, the assumed bid and offer prices are shown in Table \ref{table_cost_ben_param_2}. 

\begin{table}
	\caption{Cost and benefit parameters of flexible grid participants in general monetary \\units \UseTextSymbol{TS1}{¤} for Scenario 2.}
	\label{table_cost_ben_param_2}
	\centering
	\begin{tabular}{|c|c||c|c|}
		\hline
		& & $c_{i}$ in \UseTextSymbol{TS1}{¤}/MW \\
		\hline
		\hline
		\multirow{10}{*}{\begin{turn}{90}\rlap{\hspace{-7pt}EPN}\end{turn}} 
		& CHP & 10 \\
		& PV  & 7 \\
		& WPP & 4 \\
		& Heat pump 1 & 12 \\
		& Heat pump 2 & 9  \\
		& Heat pump 3 & 9  \\
		& Flexible Consumer Bus 3& 12 \\
		& Flexible Consumer Bus 4& 11.5 \\
		& Flexible Consumer Bus 8& 11 \\
		& Flexible Consumer Bus 9& 10 \\
		\hline
		\multirow{9}{*}{\begin{turn}{90}\rlap{\hspace{-7pt}DHN}\end{turn}} 
		& CHP & 7 \\
		& Heat pump 1 & 4.2 \\
		& Heat pump 2 & 3.5 \\
		& Heat pump 3 & 3.2 \\
		& Flexible Consumer Node 8~  & 11  \\
		& Flexible Consumer Node 10 & 10.8 \\
		& Flexible Consumer Node 13 & 10.5 \\
		& Flexible Consumer Node 16 & 10.2 \\
		& Flexible Consumer Node 18 & 10.9 \\
		\hline
	\end{tabular}
	\renewcommand{\arraystretch}{1}
\end{table}

\subsubsection{Results and Analysis - Price Signals}

\begin{figure}
	\centering
    \footnotesize
	\input{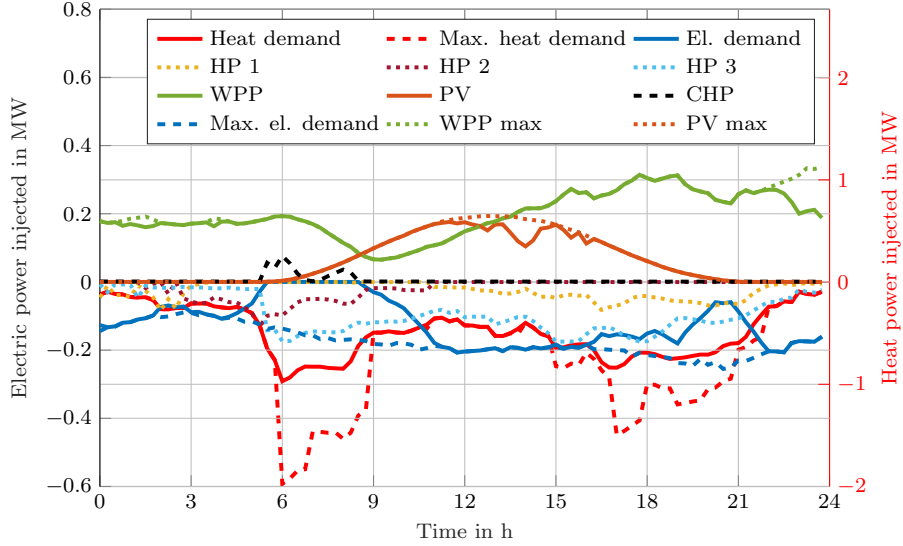}
	\caption{Injected power by all EPN participants and all DHN consumers for Scenario 2. The power injection of the CHP plant and the HPs corresponds to  electric power.}
	\label{fig_power_infeed_EPN_DHN_demand}
\end{figure}

The resulting heat and power injection and demand, are shown in Figure \ref{fig_power_infeed_EPN_DHN_demand}, where it can be seen that HPs are preferred over other flexible electric consumers by the TCS. This is particularly obvious between $\SI{5.5}{\hour}$ and $ \SI{8.5}{\hour}$ when the heat demand is high, and the RESs infeed is not sufficient to meet the entire electricity demand. Furthermore, note that the TCS utilizes HP 3 the most, as it is the highest contributor to the social welfare, while the most expensive producer, the CHP, is only run between $\SI{5.25}{\hour}$ and $ \SI{8.5}{\hour}$ to supply the peak demand.
 
\begin{figure}
	\centering
	\footnotesize
    \begin{minipage}[b]{0.55\textwidth}
    		\footnotesize
	\input{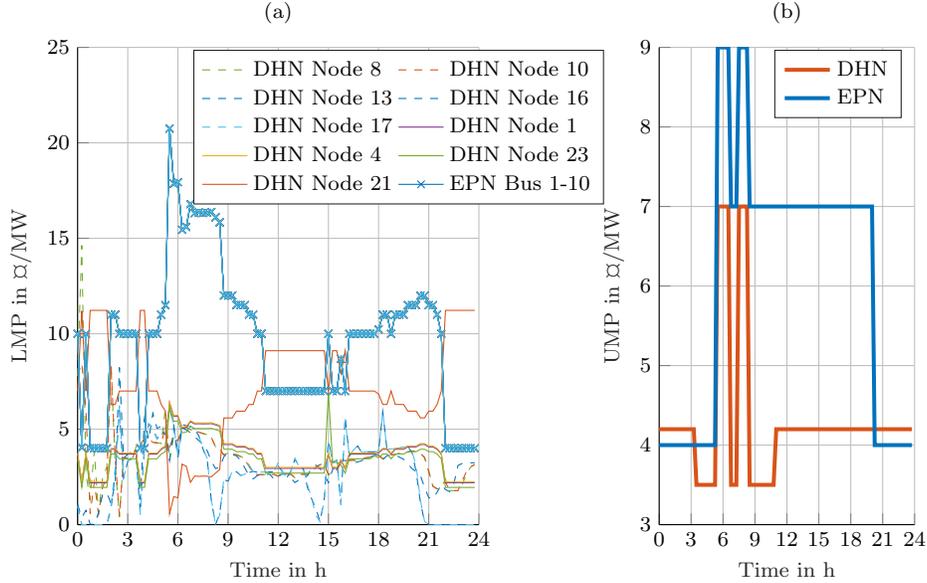}
	\end{minipage}
    \hfill
	\footnotesize
		\begin{minipage}[b]{0.35\textwidth}
				\footnotesize
%
%
\definecolor{mycolor1}{rgb}{0.00000,0.44700,0.74100}%
\definecolor{mycolor2}{rgb}{0.92900,0.69400,0.12500}%
\definecolor{mycolor3}{rgb}{0.63500,0.07800,0.18400}%
\definecolor{mycolor4}{rgb}{0.30100,0.74500,0.93300}%
\definecolor{mycolor5}{rgb}{0.46600,0.67400,0.18800}%
\definecolor{mycolor6}{rgb}{0.85000,0.32500,0.09800}%
\begin{tikzpicture}

\begin{axis}[%
title = {(b)},
at={(0.758in,0.481in)},
scale only axis,
xmin=0,
xmax=24,
xtick={ 0,  3,  6,  9, 12, 15, 18, 21, 24},
xlabel style={font=\color{white!15!black}},
xlabel={Time in h},
ymin=3,
ymax=9,
ylabel style={font=\color{white!15!black}},
ylabel={UMP in $\text{\textcurrency}$/MW},
axis background/.style={fill=white},
axis x line*=bottom,
axis y line*=left,
xmajorgrids,
ymajorgrids,
legend style={legend cell align=left, align=left, draw=white!15!black, font=\footnotesize,at={(0.7,0.995)}, anchor=north,}
]
\addplot [color=mycolor6, line width=\mylinewidth]
  table[row sep=crcr]{%
0	4.2\\
3.25	4.2\\
3.5	3.5\\
5.25	3.5\\
5.5	7\\
6.5	7\\
6.75	3.5\\
7.25	3.5\\
7.5	7\\
8.25	7\\
8.5	3.5\\
10.75	3.5\\
11	4.2\\
23.75	4.2\\
};
\addlegendentry{DHN}

\addplot [color=mycolor1, line width=\mylinewidth]
  table[row sep=crcr]{%
0	4\\
5.25	4\\
5.5	9\\
6.5	9\\
6.75	7\\
7.25	7\\
7.5	9\\
8.25	9\\
8.5	7\\
20	7\\
20.25	4\\
23.75	4\\
};
\addlegendentry{EPN}

\end{axis}
\end{tikzpicture}%
	\end{minipage}
	\caption{(a) LMPs and (b) UMPs of EPN and DHN for Scenario 2.}
	\label{fig_LMPs_UMPs}
\end{figure}

 The LMPs for this scenario are calculated from the optimization model, and the UMPs are based on \cite{MaurerGollaRichterEtAl2021}. The UMPs of the EPN and the DHN are calculated separately, based on the intersection of the dispatched offer and bid curves\footnote{If no intersection, the last accepted offer is used as the UMP.}, which are established by sorting the dispatched bids and offers by their respective prices. The resulting LMPs and UMPs\footnote{The loss costs are set to 0~$\text{\textcurrency}$/MW for the EPN and the DHN.} are shown in Figure \ref{fig_LMPs_UMPs}, respectively. Note in Figure \ref{fig_LMPs_UMPs} (a) that the LMPs show large variations, which may be attributed to some extent to the dynamics of temperature propagation throughout the network and the coupling component between the EPN and the DHN \cite{DengLiSunEtAl2019}. The UMP of the EPN in Figure \ref{fig_LMPs_UMPs} (b) shows a strong correlation with the heat power demand of the DHN, which is a logical result of the electrification of the heat supply. Furthermore, a correlation between the CHP power infeed and the UMPs of both networks can be observed, and the PV infeed depicts a weak correlation with the UMP of the EPN. Despite the well known advantages of LMPs in the context of market-based EPN operation such a pricing approach does not seem to work in this case since two different LMPs are obtained for one FNP. This results from the fact that, in DHN operation, an FNP is connected to two nodes, which can have different marginal prices, due to different constraints applied to the nodes; this happens in the context of limiting the temperature ranges of supply and return nodes. Besides, difficulties with determining LMPs based on the results of non-convex optimization problems are well known \cite{DengLiSunEtAl2019,Litvinov2010}. Therefore, the UMP-based pricing seems more appropriate in this class of operational problems. This pricing approach incentivizes the FNPs to bid with their true marginal costs in the absence of network congestions \cite{MaurerGollaRichterEtAl2021}. 
 
\subsubsection{Results and Analysis - Independent versus Coupled Operation of EPN and DHN}

In order to examine the performance of independent operation of the EPN and the DHN, in contrast to their coupled operation, five different operating cases were considered as illustrated in Table \ref{sw_cases}. For Case 3 and 4 one network was first optimized, fixing the obtained power of the energy converters to optimize the second network. These resemble the operation where either the heat or the electric power market would be cleared first, without any information exchange between the markets. In both cases, the optimization of the second network failed to converge in several time steps, as noted in Table \ref{sw_cases}. Thus, a secure network operation could not be assured in both energy networks in either Case 3 or Case 4.

A comparison of the Cases 1, 2 and 5 for their $\SI{24}{\hour}$ social welfare, shown in Table \ref{sw_cases}, shows that the joint optimization of the EPN and DHN yields the highest social welfare, whereas, if the EPN or the DHN are optimized independently, without considering the presence of the other network, the social welfare is significantly reduced. 

\begin{table}
	\renewcommand{\arraystretch}{1.1}
	\caption{Accumulated $\SI{24}{\hour}$ social welfare in various cases.}
	\label{sw_cases}
	\centering
	\begin{tabular}{|l|l|c|}
		\hline
		\multicolumn{2}{|c|}{Case } &  Social welfare in~\UseTextSymbol{TS1}{¤} \\
		\hline
		\hline
		\multirow{6}{*}{} 
		1.& EPN optimized operation, no DHN & 187.39 \\
	    	  \hline
	   ~2.& DHN optimized operation, no EPN & 321.77 \\
	    \hline
	   ~3.& EPN, DHN optimized in sequence & No Feasible Solution \\
	    	    \hline
	    ~4.& DHN, EPN optimized in sequence & No Feasible Solution \\
	    	    \hline
	    ~5.& EPN and DHN joint optimization & 443.70 \\
	     & & (EPN = 138.06; DHN = 305.65) \\
		\hline
	\end{tabular}
	\renewcommand{\arraystretch}{1}
\end{table}

\subsubsection{Results and Analysis - Varying Bid Prices} \label{sec:Res_Var_Bids}
\begin{figure}[h]   
	\centering
	\footnotesize
	\input{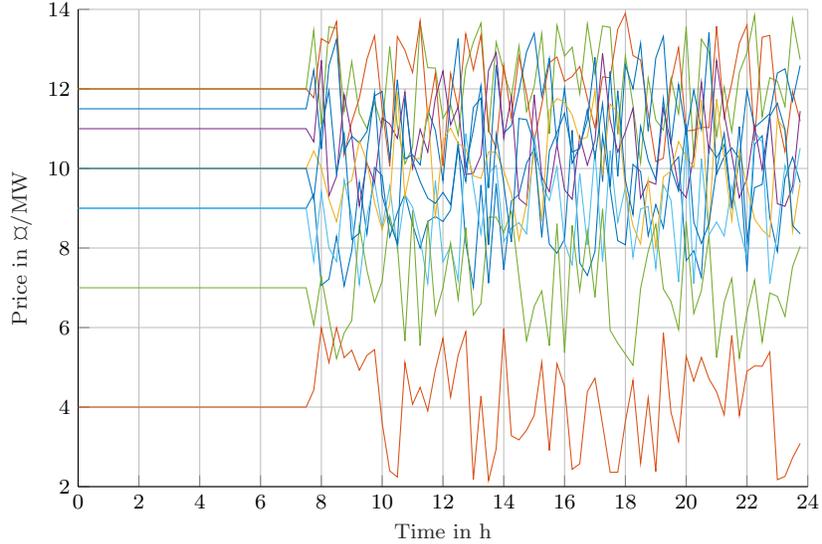}
	\caption{\textcolor{black}{Varying bid and offer prices of all FNPs in EPN for Scenario 2.}}
	\label{fig_bid_prices_szen2_3_epn}
\end{figure}

\begin{figure}[h]   
	\centering
	\footnotesize
	\input{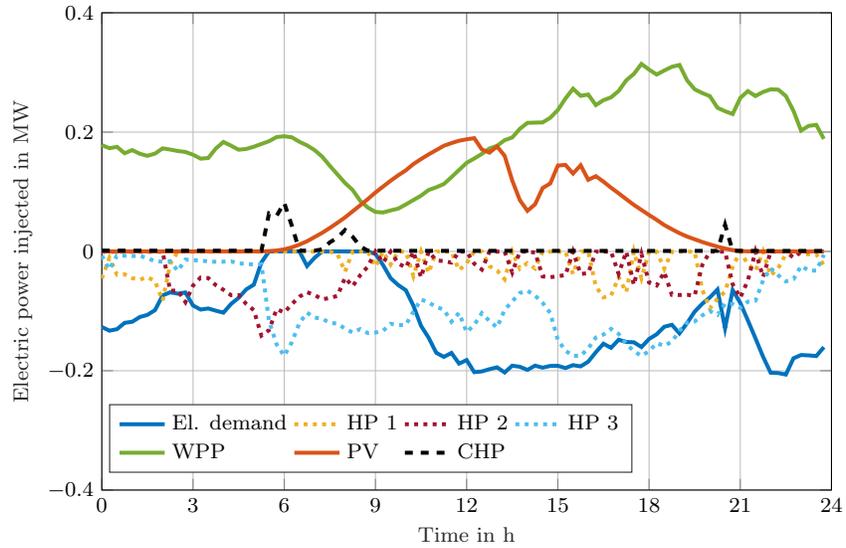}
	\caption{Electric powers of all FNPs injected into the EPN for Scenario 2 with varying bid prices.}
	\label{fig_power_infeed_EPN_szen2_3}
\end{figure}

\begin{figure}[h]
	\centering
	\footnotesize
	\begin{minipage}[b]{0.55\textwidth}
		\footnotesize
		\input{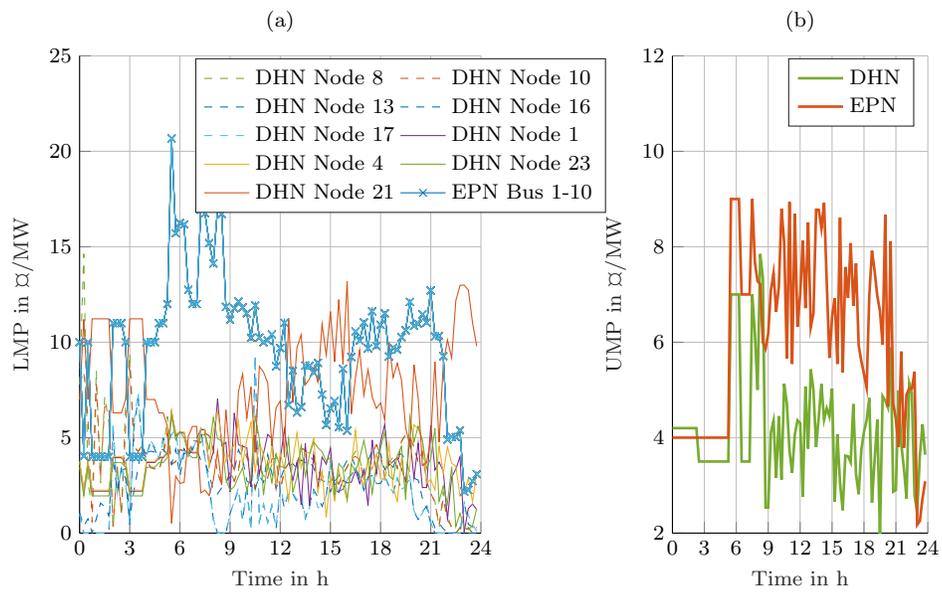}
	\end{minipage}
	\hfill
	\footnotesize
	\begin{minipage}[b]{0.35\textwidth}
		\footnotesize
%
%
\definecolor{mycolor1}{rgb}{0.46600,0.67400,0.18800}%
\definecolor{mycolor2}{rgb}{0.85000,0.32500,0.09800}%
\begin{tikzpicture}

\begin{axis}[%
	title = {(b)},
	at={(0.758in,0.481in)},
	scale only axis,
	xmin=0,
	xmax=24,
	xtick={ 0,  3,  6,  9, 12, 15, 18, 21, 24},
	xlabel style={font=\color{white!15!black}},
	xlabel={Time in h},
	ymin=2,
	ymax=12,
	ylabel style={font=\color{white!15!black}},
	ylabel={UMP in $\text{\textcurrency}$/MW},
	axis background/.style={fill=white},
	axis x line*=bottom,
	axis y line*=left,
	xmajorgrids,
	ymajorgrids,
	legend style={legend cell align=left, align=left, draw=white!15!black, font=\footnotesize,at={(0.7,0.995)}, anchor=north,}
	]
\addplot [color=mycolor1, line width=1.0pt]
  table[row sep=crcr]{%
0	4.2\\
2.25	4.2\\
2.5	3.5\\
5.25	3.5\\
5.5	7\\
6.25	7\\
6.5	3.5\\
7.25	3.5\\
7.5	7\\
7.75	6.17233032270895\\
8	5.00203351960918\\
8.25	7.84860262073945\\
8.5	7.34190984442757\\
8.75	2.53496822618317\\
9	2.5360350234743\\
9.25	4.1091493148899\\
9.5	4.35020336303069\\
9.75	3.15563173107415\\
10	4.45141872908877\\
10.25	3.11936108556556\\
10.5	5.13206824665096\\
10.75	4.71656171560577\\
11	3.73039074799223\\
11.25	3.26281395515159\\
11.5	4.06308715839256\\
11.75	4.36237966391626\\
12	4.30605602813872\\
12.25	2.76633653212809\\
12.5	5.07203593694973\\
12.75	3.77044314492661\\
13	5.43301897815281\\
13.25	5.00186968383176\\
13.5	3.80276201266745\\
13.75	4.4299584157517\\
14	5.12375365325593\\
14.25	4.37802331636928\\
14.5	4.61394035754044\\
14.75	4.44371130210954\\
15	5.02939035635543\\
15.25	2.65920833256047\\
15.5	4.07253257647828\\
15.75	3.04561328781864\\
16	2.47969160397167\\
16.25	3.62222887427465\\
16.5	3.60952947930926\\
16.75	3.48393298362719\\
17	4.70427088960772\\
17.25	3.23857082612793\\
17.5	2.81494251819304\\
17.75	3.74542006965218\\
18	4.11882267082921\\
18.25	4.36615456499474\\
18.5	4.83098581802111\\
18.75	2.63875269934897\\
19	3.53188419342846\\
19.25	4.37609793787103\\
19.5	1.96287590749218\\
19.75	4.8717151716091\\
20	4.77716229367311\\
20.25	4.63183551238401\\
20.5	5.89103785380423\\
20.75	2.86833272369914\\
21	2.90516094063521\\
21.25	4.99168843885652\\
21.5	3.87100334189109\\
21.75	3.36374349051709\\
22	2.72023587056797\\
22.25	5.14088389409748\\
22.5	4.92770369807576\\
22.75	2.78539078079925\\
23	2.44068211862032\\
23.25	3.61779250010991\\
23.5	4.28214710020445\\
23.75	3.64273215583957\\
};
\addlegendentry{DHN}

\addplot [color=mycolor2, line width=1.0pt]
  table[row sep=crcr]{%
0	4\\
5.25	4\\
5.5	9\\
6.25	9\\
6.5	7\\
7.25	7\\
7.5	9\\
7.75	7.66739205732282\\
8	7.27888861970338\\
8.25	7.21768324385895\\
8.5	5.99479772519762\\
8.75	5.86306960707304\\
9	6.17771256747573\\
9.25	7.08348347026443\\
9.5	7.44027737137406\\
9.75	6.63638107837865\\
10	7.1624798081347\\
10.25	8.79864005304418\\
10.5	8.08524966391912\\
10.75	5.6632437557724\\
11	8.93742122921389\\
11.25	5.54515653977654\\
11.5	8.68772378787027\\
11.75	6.32382963967378\\
12	6.99915470287559\\
12.25	8.12802480499939\\
12.5	6.71353709838364\\
12.75	8.51124339630632\\
13	6.31533506804539\\
13.25	6.60165780750874\\
13.5	8.77609537044884\\
13.75	8.77462792582658\\
14	8.38999745199012\\
14.25	8.92151573293766\\
14.5	7.25929781721219\\
14.75	5.67111409886873\\
15	6.53043931418366\\
15.25	6.91948233875703\\
15.5	5.54452710826556\\
15.75	8.60920832735423\\
16	5.36779118725292\\
16.25	7.57527083197485\\
16.5	6.93201296761222\\
16.75	8.07579668110504\\
17	6.76251771038799\\
17.25	7.65310887300113\\
17.5	5.93327200551405\\
17.75	5.60516309416687\\
18	5.33271491898505\\
18.25	5.05212690086537\\
18.5	6.97946759280471\\
18.75	7.91431350232994\\
19	7.49180510645983\\
19.25	6.96860467668185\\
19.5	6.65756349719249\\
19.75	5.94845161635704\\
20	8.66814940355783\\
20.25	4.65386829679575\\
20.5	8.11569907422995\\
20.75	4.71229126726831\\
21	4.38457315507917\\
21.25	3.80573480057215\\
21.5	5.80504562887094\\
21.75	3.78907146728423\\
22	4.90075438065\\
22.25	5.03399303306355\\
22.5	5.0274192823738\\
22.75	5.3833622033197\\
23	2.1743345559679\\
23.25	2.2573544979336\\
23.5	2.72514225116702\\
23.75	3.08695260870866\\
};
\addlegendentry{EPN}

\end{axis}
\end{tikzpicture}%
	\end{minipage}
	\caption{(a) LMPs and (b) UMPs of EPN and DHN for Scenario 2 with varying bid prices.}
	\label{fig_LMPs_UMPs_Szen2_3}
\end{figure}

\begin{figure}[h]   
	\centering
	\footnotesize
%
%
\definecolor{mycolor1}{rgb}{0.46600,0.67400,0.18800}%
\begin{tikzpicture}

\begin{axis}[%
at={(0.758in,0.481in)},
scale only axis,
xmin=0,
xmax=24,
xtick={ 0,  3,  6,  9, 12, 15, 18, 21, 24},
xlabel style={font=\color{white!15!black}},
xlabel={Time in h},
ymin=0,
ymax=1.4,
ylabel style={font=\color{white!15!black}},
ylabel={Calculation time in min},
axis background/.style={fill=white},
xmajorgrids,
ymajorgrids,
legend style={legend cell align=left, align=left, draw=white!15!black}
]
\addplot [color=mycolor1]
  table[row sep=crcr]{%
0	1.30286666666667\\
0.25	0.860933333333332\\
0.5	0.353899999999999\\
0.75	1.32735\\
1	0.466149999999999\\
1.25	0.262250000000002\\
1.5	0.387233333333334\\
1.75	0.394266666666667\\
2	0.432816666666668\\
2.25	0.535933333333332\\
2.5	0.479166666666668\\
2.75	0.37865\\
3	0.736466666666665\\
3.25	0.404166666666665\\
3.5	0.375266666666668\\
3.75	0.426033333333333\\
4	0.859633333333335\\
4.25	0.626300000000001\\
4.5	0.447916666666668\\
4.75	0.427866666666667\\
5	0.374749999999999\\
5.25	0.411716666666667\\
5.5	0.516400000000001\\
5.75	0.499483333333334\\
6	0.507283333333334\\
6.5	0.405733333333334\\
6.75	0.405200000000001\\
7	0.425266666666666\\
7.25	0.536983333333332\\
7.5	0.580733333333335\\
7.75	0.360150000000001\\
8	0.41405\\
8.25	0.457550000000001\\
8.5	0.372399999999999\\
8.75	0.630466666666667\\
9	0.355983333333334\\
9.25	0.412766666666666\\
9.5	0.630216666666666\\
9.75	1.1862\\
10	1.2672\\
10.25	0.622916666666665\\
10.5	0.333066666666667\\
10.75	0.483599999999999\\
11	0.359883333333332\\
11.25	0.442966666666667\\
11.5	0.327349999999999\\
11.75	0.5276\\
12	0.392183333333332\\
12.25	0.167449999999999\\
12.5	0.401033333333334\\
12.75	0.475249999999999\\
13	0.193483333333333\\
13.25	0.387766666666668\\
13.5	0.474216666666667\\
13.75	0.464583333333334\\
14	0.47345\\
14.25	0.408066666666667\\
14.5	0.299216666666666\\
14.75	0.574750000000002\\
15	0.303133333333335\\
15.5	0.491916666666668\\
15.75	0.436733333333333\\
16	0.290099999999999\\
16.25	0.472133333333332\\
16.5	0.454433333333334\\
16.75	0.374466666666667\\
17	0.407033333333334\\
17.25	0.379950000000001\\
17.5	0.645833333333332\\
17.75	0.3948\\
18	0.314050000000002\\
18.25	0.350000000000001\\
18.5	0.25\\
18.75	0.42475\\
19	0.506516666666666\\
19.25	0.548950000000001\\
19.5	0.40025\\
19.75	0.420300000000001\\
20	0.498699999999999\\
20.25	0.368483333333334\\
20.5	0.482299999999999\\
20.75	0.477083333333333\\
21	0.425249999999998\\
21.25	0.7302\\
21.5	0.402333333333335\\
21.75	0.365616666666668\\
22	0.245566666666665\\
22.25	0.310949999999998\\
22.5	0.469016666666668\\
22.75	0.45365\\
23	0.899999999999999\\
23.25	0.3474\\
23.5	0.477866666666667\\
23.75	0.393999999999998\\
};

\end{axis}
\end{tikzpicture}%
	\caption{Calculation times of the IPOPT solver for every time step in Scenario 2 with varying bid prices.}
	\label{fig_clc_time_szen2_3}
\end{figure}
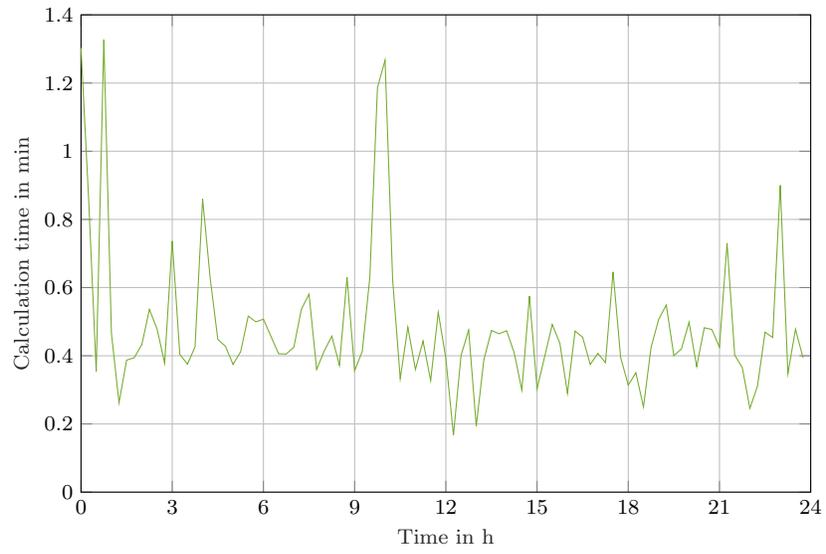

In this section, the bid prices $c_{n,k}$ are assumed to vary for all $n$ and most $k$, as depicted in \textcolor{black}{Figure \ref{fig_bid_prices_szen2_3_epn} and Figure \ref{fig_bid_prices_szen2_3_dhn}}. The resulting electric power injections of all FNPs are illustrated in Figure \ref{fig_power_infeed_EPN_szen2_3} and the corresponding price signals are shown in Figure \ref{fig_LMPs_UMPs_Szen2_3}, where the impact of the varying bidding prices on the EPN and DHN dispatch and prices can be observed. Note that the model properly captures and reflects the complex behavior of the bids on the outputs, with no significant increase in the calculation times, as depicted in Figure \ref{fig_clc_time_szen2_3}. The latter shows that the initialization in the first time step poses a high computational burden, as expected, since there is no warm start of the MPC procedure and thus all variables are far from their optimal values at the beginning of the optimization; most of the following solutions are then calculated within less than $\SI{50}{\percent}$ of the time needed for the initialization step. All calculations can be performed within $\SI{1.4}{\min}$. 

\begin{figure}[h]   
	\centering
	\footnotesize
	\input{tikzplots/szen2_3/bid_prices_dhn.tex}
	\caption{\textcolor{black}{Varying bid and offer prices of all FNPs in DHN for Scenario 2.}}
	\label{fig_bid_prices_szen2_3_dhn}
\end{figure}

\subsection{Summary \textcolor{black}{and Discussion}}

All in all, the results presented in the case studies above align with the findings in \cite{WuYanJiaEtAl2016,Mancarella2014}, which have outlined the advantages of joint operation of CEPDHNs in terms of enhanced economic and technical performance.
Note that this is the first work within the context of transactive control for CEPDHNs, wherein aspects such as different operating strategies of pumps, the effects of differential pressure regulators, the impact of loss reduction by the ISOEMS, and comparison of LMP versus UMP approaches are discussed. \newline
\textcolor{black}{Future work could focus on the following two aspects: 
The first is, that the DHN model brings along a very high modeling detail, but this detail could still be expanded to the case of varying flow directions as these occur in DHNs with meshed supply networks. A second aspect that can be researched in future work is the fact that the calculation time of the optimization problem of the ISOEMS will most likely not be solvable in real time for CEPDHNs spanning large regions, e.g. with multiple hundreds of nodes. Thus, distributed optimization approaches could be used here to enable a parallel and thereby faster computation of multiple ISOEMSs.}

\FloatBarrier

\section{Conclusions}

This paper presented the design of a new TCS for an optimal market- and control-based operation of an integrated CEPDHN multi-energy system. This is a novel application of TC to such systems, as state-of-the-art TE approaches for EPNs cannot be simply transferred to DHNs and CEPDHNs, due to the fundamental operational differences between the two networks. The TCS was designed at two hierarchy levels, comprising the EMSSAs on the first level and the ISOEMS on the second level. The ISOEMS was built using an iterative approach where the auxiliary values of the DHN model were recalculated after every optimization, and used in the next time step. This reduced the complexity of the optimization problem while still considering the heat propagation through pipelines, based on a close approximation of the node method with variable temperatures and mass flows. The presented ISOEMS design model also incorporated a detailed hydraulic model considering the pressure differences and mass flows caused by varying consumer behavior and actively controlled hydraulic components, which were employed for efficient DHN operation with low losses. Additionally, the presented approach guaranteed a high level of privacy to the FNPs, as only the bids and offers from the EMSSAs were communicated to the ISOEMS. Realistic case studies showed the effectiveness of the proposed TCS, which enabled optimal provision of flexibility through DSM in the power system and the DHN as well as through storing thermal energy in the pipelines of the DHN. 

\section*{Funding} 

This work was supported by DAAD through KHYS, KIT; and NSERC Canada.~

\section{Appendix A}

This appendix illustrates the accuracy of the thermal pipeline model presented in Section \ref{sec_DHN_model}. This model is based on the well-known Node Method \cite{Benonysson1991}, which has been shown to be precise in \cite{Benonysson1991,MaurerRaRatzelMalanEtAl2021,Palsson2000,GabrielaitieneSundenKacianauskasEtAl2003}, as demonstrated here for exact mass flow predictions. Comparisons of measurement data with this method showed small differences in RMSEs temperature values between \SI{0.507}{\kelvin} and \SI{1.038}{\kelvin} in [35]. Furthermore, as stated in \cite{SunWangWuEtAl2019}, the model allows to obtain high quality mass flow predictions for DHNs, with maximum errors in the mass flow mean values lower than \SI{1}{\percent} and maximum standard deviation errors lower than \SI{6}{\percent}. 

\begin{figure}[h]   
	\centering
	\footnotesize
	\input{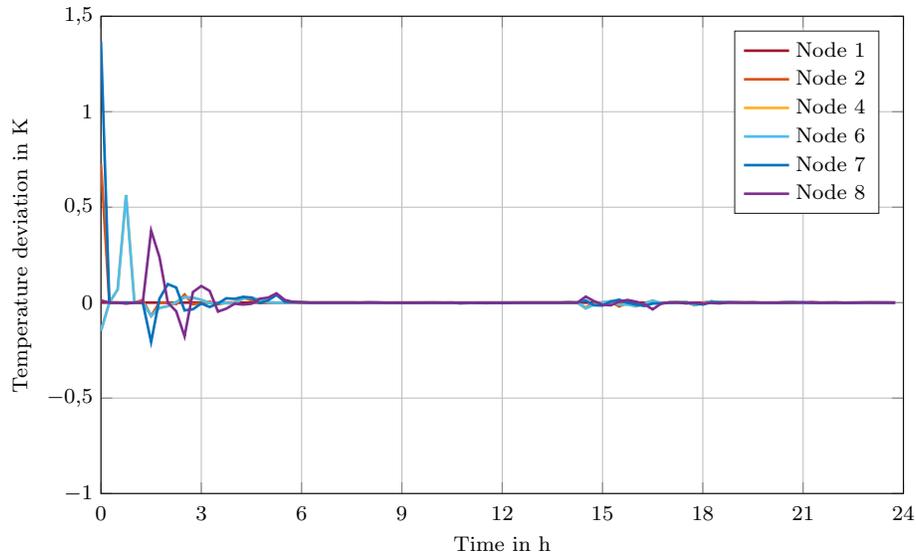}
	\caption{Deviation between the exact simulated temperature and the one calculated from the proposed approximation}
	\label{fig_dev_approx_node}
\end{figure}

The model for a DHN pipeline proposed in \cite{Benonysson1991}, enhanced by adding (\ref{eq_verweildauer}) to account for the length of stay of the water mass in the pipelines, is compared here with the approximate model proposed in the paper and using parameters taken from \cite{Oppelt2015,EnergieSchweiz2018}. The results of this comparison are shown in Figure \ref{fig_dev_approx_node}, where it can be observed that the maximum temperature deviation is about \SI{1.4}{\kelvin}, which decreases in time to less than \SI{0.1}{\kelvin}. This demonstrates that the proposed model is highly accurate.


\bibliography{Quellen_PmN}

\end{document}